\documentclass[12pt]{article}
\usepackage[utf8]{inputenc}
\usepackage[russian,ukrainian]{babel}
\usepackage[OT1]{fontenc}
\usepackage{amsmath}
\usepackage{amsfonts}
\usepackage{amssymb}
\usepackage{graphicx}
\usepackage[left=2cm,right=2cm,top=2cm,bottom=2cm]{geometry}
\usepackage[numbers,sort&compress]{natbib}
\usepackage[colorlinks=true,citecolor=blue,filecolor=magenta,unicode]{hyperref}

\title{Тричастинкові поля як метод опису баріонів у процесах розсіяння}

\author{Потієнко О.С., Шарф І.В., Зеленцова Т.М., Чудак Н.О., Небога Г.Г., \\ 
Меркотан К.К., Пташинський Д.А.}
\date{05.11.2020}

\begin{document}

\maketitle

\begin{center}
\begin{minipage}{15cm}
 \textit{Одеський національний політехнічний університет, \\
     \mbox{~~}проспект Шевченка 1, Одеса, 65000, Україна \\ 
}     
\end{minipage}
\end{center}

\begin{abstract}
В роботі ми пропонуємо модель, у межах якої одночасно можна описати утримання кварків усередині протону і взаємодію цих кварків із кварками іншого адрону. Для цього використовується модель тричастинкового біспінорного поля. Таке поле на відміну від звичайного одночастинкового поля розглядається не на просторі Мінковського, а на підмножині тензорного добутку трьох просторів Мінковського, яка характеризується рівністю часових координат трьох подій (підмножина одночасності). Таку підмножину неможливо виділити Лоренц інваріантним способом. Але внаслідок закону перетворення багаточастинкового поля, при переході від однієї інерційної системи відліку до іншої, динамічні рівняння для тричастинкового поля узгоджуються із принципом відносності. Взаємодія між кварками вводиться шляхом вимоги локальної $ SU_{c}\left( 3\right) $ інваріантності. Але в роботі запропоновано більш загальний спосіб досягнення цієї інваріантності ніж заміна в лагранжіані звичайних похідних на коваріантні. Застосована процедура приводить до двоіндексного тензорного калібрувального поля, як по лоренцевим, так і по внутрішнім індексам. Обговорюється зв'язок цього поля із зв'язаними станами глюонів - глюболами. Показано, що це глюбольне поле може бути представлено в виді суми двох доданків: класичного поля, яке грає роль потенційної енергії взаємодії кварків у протоні і поля, що відповідає квантовим флуктуаціям навколо класичного поля, і описує взаємодію кварків протона з іншими адронами. Для поля, що визначає потенційну енергію взаємодії кварків отримано динамічне рівняння. Показано, що це рівняння має розв'язки, які відповідають конфайнменту кварків. 

\end{abstract}

\section{Вступ}
На наш погляд, у квантовій теорії поля існує принципова проблема. Суть цієї проблеми полягає в тому, що оператори, за допомогою яких можна побудувати фоківський стан системи взаємодіючих частинок \cite{Landau_Paijerls_prostir_Foca, Foc_prostir_Foca, Berezin_prostir_Foca,Bogolyubov_rus} відрізняються від операторів квантованих полів. Так відбувається тому, що оператори польових функцій визначені на просторі Мінковського і, відтак, у довільній системі відліку залежать від часу. Внаслідок цього, такі оператори, діючи на фоківський стан змінюють не тільки числа заповнення одночастинкових станів, але й часову залежність стану. І ця зміна відбувається незалежно від оператора часової еволюції. В той же час, для системи взаємодіючих частинок, залежність стану від часу не може бути зведена до залежностей від часу одночастинкових станів, і визначається виключно оператором часової еволюції. Визначення чисел заповнення одночастинкових станів, у цьому випадку, потребуватиме відділення часової залежності від залежностей решти динамічних змінних, з подальшим розкладом цієї останньої залежності по добутках одночастинкових станів \cite{Bogolubov_vtorinne_kvantuvanna}. Таким чином, ми отримаємо оператори, дія яких змінює залежність стану від чисел заповнення, але не змінює безпосередньо залежність стану від часу. Тобто, вони змінюють залежність стану від чисел заповнення, а вже як наслідок зміни цієї залежності, діючи на такий змінений стан оператором часової еволюції, ми отримаємо зміну його залежності від часу. Тим самим, вони відрізняються від утворюючих елементів алгебри польових операторів і незрозуміло за якими правилами ці оператори повинні переставлятися. Тому, дія польових операторів і оператора часової еволюції на елементи простору Фока системи взаємодіючих частинок стає невизначеною. Цю ситуацію можна добре прослідкувати на прикладі відомого методу Тамма-Данкова \cite{Tamm_Dancoff_Tamm1945,Tamm_Dancoff_DancoffRev.78.382,Tamm_Dancoff_Silin:1955,Light_front_Tamm_Dancoff}. В роботі \cite{Tamm_Dancoff_Tamm1945} розглянута нами проблема ігнорується внаслідок того, що ігнорується як залежність від часу фоківського стану, так і залежність від часу польових операторів. Тоді, оператори, за допомогою яких виражається стан і оператори, що входять в гамільтоніан в обох випадках виявляються функціями від імпульсу. В роботі \cite{Tamm_Dancoff_Tamm1945} їх переставляють за одночасними переставними співвідношеннями \cite{Bogolyubov_rus}, що формально дозволяє визначити дію гамільтоніану на стан і розв'язувати для нього задачу на власні значення. Однак, на наш погляд, такий підхід є помилковим, тому що оператори народження, за допомогою яких будується стан, народжують частинку з певним тривимірним імпульсом, тобто збільшують відповідне число заповнення в розкладі координатної частини стану по одночастинкових станах, власних для імпульсу. Польові оператори народження, які в представленні взаємодії також залежать від тривимірного імпульсу, народжують частинку не тільки з таким імпульсом, а ще й з певною енергією, яка повністю визначається цим тривимірним імпульсом. Тому, це є різні операторнозначні функції від тривимірного імпульсу і переставляти їх за переставними співвідношеннями для операторів одного й того самого поля не має підстав. Взагалі, будь-який польовий оператор, виходячи з координатного представлення, якщо застосувати для нього перетворення Фур'є, буде народжувати або знищувати частинку не тільки з якимось імпульсом, а й із якоюсь енергією. Це добре видно в роботі \cite{Morgan:2015hza}, де багаточастинкові ефекти розглядаються за допомогою звичайних одночастинкових польових операторів.  Зазначені проблеми проявляються в роботах \cite{Tamm_Dancoff_DancoffRev.78.382,Light_front_Tamm_Dancoff}, в яких метод Тамма-Данкова намагаються сформулювати Лоренц-коваріантним чином. Це миттєво призводить до розгляду багаточасових амплітуд ймовірності. Нашу точку зору щодо неможливості застосування багаточасового опису в релятивістській квантовій теорії ми докладно виклали в роботі \cite{Ptashynskiy2019MultiparticleFO}.

В тих задачах, в яких початковий і кінцевий стан можна сформувати з вільних квантів взаємодіючих полів, як наприклад, у задачах розсіяння лептонів, вказана проблема не виникає, бо якщо частинки не взаємодіють, їх одночастинкові стани можна зсувати в часі поодинці, незалежно один від одного, і тоді такий стан можна виразити, діючи польовими операторами на вакуум і переставляючи далі із тими ж польовими операторами, що входять в оператор часової еволюції. При цьому, матимемо динамічну модель, що призведе до збереження суми одночастинкових енергій-імпульсів, що в цьому випадку відповідає експерименту. 

Але в задачах розсіяння адронів, кварки всередині адронів сильно взаємодіють як в початковому стані, так і в кінцевому. Тому, розглянута проблема стає суттєвою. На сьогоднішній день при описі процесів з адронами, цю проблему обходять розділяючи процес розсіяння на стадії \cite{Bassetto:1984ik, Sj_strand_2006}, і описуючи кожну стадію окремо за допомогою розробленого спеціально для неї підходу. Зокрема, вихідні адрони, зазвичай, представляють за допомогою партонної моделі \cite{Feynman:1973xc}. Такий підхід не може вважатися задовільним розв'язком вказаної проблеми. По-перше, адрон замінюється системою невзаємодіючих частинок і, таким чином, в задачу штучно вводяться одночастинкові енергії-імпульси. По-друге, партонна модель оперує не амплітудами, а ймовірностями \cite{Dokshitzer:1977sg,Altarelli:1977zs} і тому такої конструкції як оператор народження партону, або лагранжіан партонного поля, в межах цієї моделі, не може бути в принципі. Тому, партнонна модель це скоріше спосіб феноменологічно обійти розглянуту проблему, ніж її розв'язати. Те ж саме можна сказати щодо стадії адронізації \cite{Bassetto:1984ik, Sj_strand_2006}. Оскільки, оператор взаємодії полів містить дельта-функцію, що забезпечує збереження сум одночастинкових енергій-імпульсів \cite{CARAVAGLIOS,DRAGGIOTIS1998157}, то як наслідок, і оператор часової еволюції буде забезпечувати лише зберігання суми одночастинкових енергій-імпульсів, в той час як для адронів не існує одночастинкових енергій-імпульсів і в експерименті зберігається лише сумарний чотиривектор енергії-імпульсу адронів. Цей висновок, мабуть, не залежить від методу розрахунку оператора часової еволюції. Дійсно, часова залежність виробляючого функціонала цього оператора \cite{Berezin_prostir_Foca} з'являється лише через його одночастинкові аргументи \cite{Faddeev_Slavnov}. Якщо застосувати для цих аргументів перетворення Фур'є, і перейти від оператора часової еволюції до оператора розсіяння, то ядра оператора розсіяння, що описують відображення між підпросторами простору Фока, які відповідають різним частинкам \cite{Berezin_prostir_Foca} будуть містити дельта-функції від різниці сум одночастинкових енергій-імпульсів у початковому і кінцевому станах. Тому, навіть, якщо б вдалося точно розрахувати континуальний інтеграл, що виражає оператор часової еволюції, мабуть, це не допомогло б описати адронізацію кварків і глюонів. З цієї точки зору зрозуміло, що в цьому сенсі не допоможуть ґраткові розрахунки \cite{PhysRevD.10.2445,RevModPhys.50.561,Wilson1977,CreutzKvarci_Gluoni}. Фактично, такі розрахунки є способом наближеного розрахунку континуального інтегралу, за допомогою якого виражається оператор розсіяння, пов'язаний із амплітудою переходу з одного стану невзаємодіючих частинок до іншого стану також невзаємодіючих частинок.     

В роботах \cite{Korotca_statta_v_UJP, Ptashynskiy2019MultiparticleFO} було запропоновано метод багаточастинкових полів для опису двочастинкових зв'язаних станів - мезонів і зв'язаних станів калібрувальних бозонів. В цих роботах, а також в \cite{Chudak:2016, Volkotrub:2015laa} було наведене мотивування цього методу і проаналізовані відмінності від інших підходів до опису зв'язаних станів кварків і глюонів у процесах розсіяння адронів. Опис експериментів із розсіяння, в яких у початковому стані використовуються розігнані протони, потребує побудови аналогічної моделі для трикваркових систем. Подібна модель вже розглядалася в роботі \cite{Chudak:2016} і навіть застосовувалася в роботі \cite{Prujne_rozsijanna_protoniv_JFS} для опису пружного розсіяння протонів. Однак, у згаданих роботах \cite{Korotca_statta_v_UJP, Ptashynskiy2019MultiparticleFO} було запропоновано більш послідовний варіант методу багаточастинкових полів. Метою цієї роботи є опис трикваркової системи в межах цього підходу. В усіх подальших обчисленнях використовується система одиниць, у якій дія і гранична швидкість передачі взаємодії безрозмірні, і всі величини домножуються на такі комбінації константи Планка $\hbar$ і граничної швидкості $c$, щоб отримати ступені довжини.

\section{Підмножина одночасності і скалярний добуток на цій підмножині}
Розглянемо спочатку систему трьох невзаємодіючих ферміонів, а потім врахуємо взаємодію між ними, яка призводить до утворення зв'язаного стану. Всі події, які можуть відбуватися з такою системою описуються тензорним добутком трьох просторів Мінковського. Кожну точку такого тензорного добутку можна охарактеризувати за допомогою дев'ятивимірного стовпця виду: 
\begin{equation}\label{9Dstovpec}
{{z}^{a}}=\left( \begin{matrix}
x_{1}^{0}  \\
x_{1}^{1}  \\
x_{1}^{2}  \\
x_{1}^{3}  \\
x_{2}^{0}  \\
x_{2}^{1}  \\
x_{2}^{2}  \\
x_{2}^{3}  \\
x_{3}^{0}  \\
x_{3}^{1}  \\
x_{3}^{2}  \\
x_{3}^{3}  \\
\end{matrix} \right).
\end{equation} 
Тут нижні індекси означають номера частинок, а верхні - відповідні часові або просторові координати у просторі Мінковського кожної частинки. Той факт, що частинки тотожні і насправді не мають номерів можна врахувати, як зазвичай, накладаючи відповідні умови симетрії на залежності від цих координат. Розглянемо лінійний простір стовпців \eqref{9Dstovpec} і введемо на ньому скалярний добуток:
\begin{equation}\label{Skalarnij_dobutoc}
\left\langle  z | z \right\rangle =\frac{1}{3}\left( g_{{{a}_{1}}{{a}_{2}}}^{Minc}x_{1}^{{{a}_{1}}}x_{1}^{{{a}_{2}}}+g_{{{a}_{1}}{{a}_{2}}}^{Minc}x_{2}^{{{a}_{1}}}x_{2}^{{{a}_{2}}}+g_{{{a}_{1}}{{a}_{2}}}^{Minc}x_{3}^{{{a}_{1}}}x_{3}^{{{a}_{2}}} \right). 
\end{equation}
Тут $ g_{{{a}_{1}}{{a}_{2}}}^{Minc}-$ тензор Мінковського. Далі, зручно буде розглядати цей лінійний простір у координатах Якобі:
\begin{equation}\label{CoordinatiJacobi}
\begin{split}
& x_{1}^{a}={{X}^{a}}-\frac{1}{3}y_{1}^{a}-\frac{1}{2}y_{2}^{a} \\ 
& x_{2}^{a}={{X}^{a}}-\frac{1}{3}y_{1}^{a}+\frac{1}{2}y_{2}^{a} \\ 
& x_{3}^{a}={{X}^{a}}+\frac{2}{3}y_{1}^{a} \\ 
\end{split}
\end{equation}
В цих координатах скалярний добуток \eqref{Skalarnij_dobutoc} приймає вид:
\begin{equation}\label{Scalarnij_dobutok_v_koordinatax_Jacobi}
\left\langle  z | z \right\rangle =g_{{{a}_{1}}{{a}_{2}}}^{Minc}\left( {{X}^{{{a}_{1}}}}{{X}^{{{a}_{2}}}}+\frac{2}{9}y_{1}^{{{a}_{1}}}y_{1}^{{{a}_{2}}}+\frac{1}{6}y_{2}^{{{a}_{1}}}y_{2}^{{{a}_{2}}} \right).
\end{equation}
Квантовий стан системи, що розглядається, описуватиметься стовпцем простору Фока, у якого відмінна від нуля лише тричастинкова компонента. Квадрат модуля цієї компоненти має сенс сумісної густини ймовірностей результатів вимірювання динамічних змінних трьох частинок, проведеного одночасно відносно тієї системи відліку, відносно якої проводиться вимірювання. Принциповість питання щодо одночасності докладно обговорена в роботах \cite{Ptashynskiy2019MultiparticleFO,Chudak_2016Internal_states}. Таким чином, фоківський стан розглядається не на лінійному просторі стовпців \eqref{9Dstovpec}, а на його підмножині, що визначається співвідношеннями:
\begin{equation}\label{Pidmnojina_odnochasnosti}
y_{1}^{0}=0,y_{2}^{0}=0,
\end{equation}
які називатимемо далі підмножиною одночасності. Також введемо позначення: 
\begin{equation}\label{Vnutrihni_coordinsti}
{{\vec{y}}_{1}}=\left( y_{1}^{1},y_{1}^{2},y_{1}^{3} \right),{{\vec{y}}_{2}}=\left( y_{2}^{1},y_{2}^{2},y_{2}^{3} \right). 
\end{equation}
Множину цих величин будемо далі називати внутрішніми координатами динамічної системи, що розглядається. Тричастинкову компоненту фоківського стовпця позначимо ${{\Psi }_{3}}\left( X,{{{\vec{y}}}_{1}},{{{\vec{y}}}_{2}} \right)$. Тут ми позначили як $X$ четвірку чисел $ {{X}^{0}},{{X}^{1}},{{X}^{2}},{{X}^{3}}$.

Точку підмножини одночасності можна охарактеризувати десятивимірним стовпцем:
\begin{equation}\label{Stovpec_na_pidmnojini_odnochasnosti}
	q=\left( \begin{matrix}
		{{X}^{0}}  \\
		{{X}^{1}}  \\
		{{X}^{2}}  \\
		{{X}^{3}}  \\
		y_{1}^{1}  \\
		y_{1}^{2}  \\
		y_{1}^{3}  \\
		y_{2}^{1}  \\
		y_{2}^{2}  \\
		y_{2}^{3}  \\
	\end{matrix} \right).
\end{equation}

Для таких стовпців введемо скалярний добуток так, щоб він співпадав із \eqref{Scalarnij_dobutok_v_koordinatax_Jacobi} у випадку, коли виконується умова одночасності \eqref{Pidmnojina_odnochasnosti}.
\begin{equation}\label{Scalarnij_dobutoc_na_pidmnojini_odnochasnosti}
	\left\langle  q | q \right\rangle ={{g}_{{{a}_{1}}{{a}_{2}}}}{{q}^{{{a}_{1}}}}{{q}^{{{a}_{2}}}},{{g}_{{{a}_{1}}{{a}_{2}}}}=\left( \begin{matrix}
		1 & 0 & 0 & 0 & 0 & 0 & 0 & 0 & 0 & 0  \\
		0 & -1 & 0 & 0 & 0 & 0 & 0 & 0 & 0 & 0  \\
		0 & 0 & -1 & 0 & 0 & 0 & 0 & 0 & 0 & 0  \\
		0 & 0 & 0 & -1 & 0 & 0 & 0 & 0 & 0 & 0  \\
		0 & 0 & 0 & 0 & -{9}/{2}\; & 0 & 0 & 0 & 0 & 0  \\
		0 & 0 & 0 & 0 & 0 & -{9}/{2}\; & 0 & 0 & 0 & 0  \\
		0 & 0 & 0 & 0 & 0 & 0 & -{9}/{2}\; & 0 & 0 & 0  \\
		0 & 0 & 0 & 0 & 0 & 0 & 0 & -6 & 0 & 0  \\
		0 & 0 & 0 & 0 & 0 & 0 & 0 & 0 & -6 & 0  \\
		0 & 0 & 0 & 0 & 0 & 0 & 0 & 0 & 0 & -6  \\
	\end{matrix} \right)
\end{equation} 

Звичайно, підмножину одночасності неможливо виділити Лоренц-інваріантним способом і у кожного інерційного спостерігача така підмножина своя. Як докладно показано в \cite{Ptashynskiy2019MultiparticleFO} це не протирічить принципам теорії відносності. Так відбувається тому, що виходячи із принципу відносності, якщо один інерційний спостерігач проводячи вимірювання над певною квантовою системою повинен робити їх одночасно відносно себе, то й кожен інший інерційний спостерігач повинен робити свої вимірювання одночасно відносно себе. Тому, різні інерційні спостерігачі не можуть скористатися для своїх вимірювань однією й тією ж системою подій. Як відомо, перетворення Лоренца пов'язують координати однієї і тієї ж події в різних інерційних системах відліку. Оскільки, при вимірюваннях різні спостерігачі повинні реалізовувати різні події, то й координати цих подій не повинні пов'язуватися перетвореннями Лоренца. Також в \cite{Chudak_2016Internal_states,Ptashynskiy2019MultiparticleFO} показано, що залежність фоківського стану від внутрішніх змінних залишається однаковою в різних інерційних системах відліку, а залежність від координат $ {{X}^{0}},{{X}^{1}},{{X}^{2}},{{X}^{3}}$ перетворюється при переході від однієї інерційної системи відліку до іншої по закону звичайної скалярної функції:
\begin{equation}\label{Scalarnij_zacon}
{{{\Psi }'}_{3}}\left( {X}',{{{\vec{y}}}_{1}},{{{\vec{y}}}_{2}} \right)={{\Psi }_{3}}\left( X={{{\hat{\Lambda }}}^{-1}}\left( {{X}'} \right),{{{\vec{y}}}_{1}},{{{\vec{y}}}_{2}} \right). 
\end{equation}
Тут $ {{{\Psi' }}_{3}}\left( {X'},{{{\vec{y}}}_{1}},{{{\vec{y}}}_{2}} \right)- $тричастинкова компонента фоківського стану в системі відліку, яка отримується від вихідної системи відліку, відносно якої тричастинкова компонента є ${{\Psi }_{3}}\left( X,{{{\vec{y}}}_{1}},{{{\vec{y}}}_{2}} \right)$ за допомогою перетворення Лоренца $\hat{\Lambda }$. Звернемо увагу на те, що зв'язок $ X={{\hat{\Lambda }}^{-1}}\left( {{X'}} \right) $ з'являється в \eqref{Scalarnij_zacon} не за рахунок перетворень Лоренца, а за рахунок перетворення форми залежності функції ${{\Psi }_{3}}\left( X,{{{\vec{y}}}_{1}},{{{\vec{y}}}_{2}} \right)$ від аргументів $ {{X}^{0}},{{X}^{1}},{{X}^{2}},{{X}^{3}}$ \cite{Bogolyubov_rus,Chudak_2016Internal_states}. Пояснимо це більш докладно. Припустимо, що в нас є два інерційні спостерігачі, які користуються системами відліку <<без штриха>> і <<з штрихом>>. Припустимо, що перший із них, використовуючи свій ансамбль тричастинкових систем, в різні моменти часу ${{X}^{0}}$ провів вимірювання координат трьох частинок ${{\vec{x}}_{1}}=\left( x_{1}^{1},x_{1}^{2},x_{1}^{3} \right),{{\vec{x}}_{2}}=\left( x_{2}^{1},x_{2}^{2},x_{2}^{3} \right),{{\vec{x}}_{3}}=\left( x_{3}^{1},x_{3}^{2},x_{3}^{3} \right)$ кожен раз одночасно відносно себе, і за результатами цих вимірювань розрахував розташування центру мас тричастинкової системи за допомогою вектору $\vec{X}=\left( {{X}^{1}},{{X}^{2}},{{X}^{3}} \right)$, і внутрішні координати ${{\vec{y}}_{1}},{{\vec{y}}_{2}}$ за формулами \eqref{CoordinatiJacobi}. Припустимо, що те ж саме зробив і спостерігач <<з штрихом>>, але з своїм, таким самим як і у першого спостерігача ансамблем таких самих тричастинкових систем. Як того вимагає принцип відносності, він реалізовував аналогічні вимірювання одночасно відносно себе. Моменти часу, коли він це робив позначимо ${{{X}'}^{0}}$, а результати вимірювань ${{{\vec{x'}}}_{1}}=\left( {x'}_{1}^{1},{x'}_{1}^{2},{x'}_{1}^{3} \right),{{{\vec{x'}}}_{2}}=\left( {x'}_{2}^{1},{x'}_{2}^{2},{x'}_{2}^{3} \right),{{{\vec{x'}}}_{3}}=\left( {x'}_{3}^{1},{x'}_{3}^{2},{x'}_{3}^{3} \right)$ і, відповідно,  ${\vec{X'}}=\left( {{{{X'}}}^{1}},{{{{X'}}}^{2}},{{{{X'}}}^{3}} \right)$ і ${{{\vec{y'}}}_{1}}=\left( {y'}_{1}^{1},{y'}_{1}^{2},{y'}_{1}^{3} \right),{{{\vec{y'}}}_{2}}=\left( {y'}_{2}^{1},{y'}_{2}^{2},{y'}_{2}^{3} \right)$. Оскільки події, які полягають в тому, що перший спостерігач виявляє частинки в момент часу ${{X}^{0}}$ в околах точок ${{\vec{x}}_{1}},{{\vec{x}}_{2}},{{\vec{x}}_{3}}$ є одночасні в одній системі відліку, а події, що полягають у виявленні другим спостерігачем у момент часу ${{{X'}}^{0}}$ частинок в околах точок ${{{\vec{x'}}}_{1}},{{{\vec{x'}}}_{2}},{{{\vec{x'}}}_{3}}$ одночасні в іншій системі відліку, то це різні трійки подій. Тому, їх координати неможливо пов'язати між собою. Але аргументи, наведені в \cite{Chudak_2016Internal_states,Ptashynskiy2019MultiparticleFO} призводять до висновку, що на кожну реалізацію вимірювання спостерігачем <<без штриха>> в момент ${{X}^{0}}$ з результатом в околі $\left( \vec{X},{{{\vec{y}}}_{1}},{{{\vec{y}}}_{2}} \right)$ знаходиться реалізація вимірювання спостерігачем <<з штрихом>> в момент ${{{X'}}^{0}}$ з результатом в околі $\left( {\vec{X'}},{{{\vec{y}}}_{1}},{{{\vec{y}}}_{2}} \right)$, такому, що величини $\left( {{{{X'}}}^{0}},{{{{X'}}}^{1}},{{{{X'}}}^{2}},{{{{X'}}}^{3}} \right)$ пов'язані із $ {{X}^{0}},{{X}^{1}},{{X}^{2}},{{X}^{3}} $ співвідношенням $ {{{X'}}^{a}}=\Lambda _{b}^{a}{{X}^{b}}, $ де $ \Lambda _{b}^{a} -$елементи матриці перетворення Лоренца, що переводить систему відліку <<без штриха>> в систему з <<з штрихом>>. Саме такий сенс вкладається в формулу \eqref{Scalarnij_zacon} і саме це ми маємо на увазі, коли говоримо, що зв'язок між координатами $X$ виникає не за рахунок перетворень Лоренца, а за рахунок перетворення форми стану \cite{Bogolyubov_rus,Chudak_2016Internal_states}.

Зазначена в попередньому розділі проблема виникає внаслідок того, що польові оператори визначені на просторі Мінковського, в той час, як фоківський стан, на який вони повинні діяти, визначений на підмножині одночасності. На нашу думку, шляхом виходу з цієї ситуації є побудова польової моделі з операторами поля, що задані не на просторі Мінковського, а на підмножині одночасності. Такі поля будемо далі називати багаточастинковими полями. Польові оператори таких багаточастинкових полів змінюватимуть числа заповнення не одночастинкових, а багаточастинкових (в нашому випадку тричастинкових) станів. Розглянемо побудову такої моделі.

\section{Тричастинкове біспінорне поле на підмножині одночасності}
Підмножина одночасності із визначеним скалярним добутком є областю визначення тричастинкового поля. Розглянемо область значень такого поля. Нас цікавить побудова операторів, які змінюватимуть числа заповнення трикваркових станів. Для випадку невзаємодіючих кварків, такий оператор можна отримати перемножаючи три оператори біспінорного поля з подальшим переходом на підмножину одночасності:
\begin{equation}\label{Trschastincove_bispinorne_pole}
{{\left. {{{\hat{\Psi }}}_{{{s}_{1}}{{s}_{2}}{{s}_{3}},{{f}_{1}},{{f}_{2}},{{f}_{3}},{{c}_{1}},{{c}_{2}},{{c}_{3}}}}\left( {{x}_{1}},{{x}_{2}},{{x}_{3}} \right) \right|}_{x_{1}^{0}=x_{2}^{0}=x_{3}^{0}}}={{\left. {{{\hat{\Psi }}}_{{{s}_{1}},{{f}_{1}},{{c}_{1}}}}\left( {{x}_{1}} \right){{{\hat{\Psi }}}_{{{s}_{2}},{{f}_{2}},{{c}_{2}}}}\left( {{x}_{2}} \right){{{\hat{\Psi }}}_{{{s}_{3}},{{f}_{3}},{{c}_{3}}}}\left( {{x}_{3}} \right) \right|}_{x_{1}^{0}=x_{2}^{0}=x_{3}^{0}}}
\end{equation}
Тут ${{s}_{1}},{{s}_{2}},{{s}_{3}}-$біспінорні індекси, ${{f}_{1}},{{f}_{2}},{{f}_{3}}-$ароматові індекси, ${{c}_{1}},{{c}_{2}},{{c}_{3}}-$кольорові індекси. Отже, область значень тричастинкового поля - лінійний простір тензорів, на якому реалізуються тензорний добуток трьох біспінорних представлень групи Лоренца, тензорний добуток трьох фундаментальних представлень ароматової і кольорової груп $SU\left( 3 \right).$ Оскільки, ми далі збираємось враховувати лише сильну взаємодію між кварками, яка призводить до їх конфайнменту і існування адрону як зв'язаного стану трьох кварків, ми ігноруватимемо ароматову структуру тензора тричастинкового поля. Тому, розглядаючи тензори далі, ми не будемо виписувати ароматові індекси. Ми хочемо побудувати оператори, які б змінювали числа заповнення протонних тричастинкових станів, тобто таких, що відповідають частинкам із спіном ${1}/{2}\;.$ Тому, з лінійного простору тензорів ${{\hat{\Psi }}_{{{s}_{1}}{{s}_{2}}{{s}_{3}}}}$ можна виділити інваріантний підпростір, на якому реалізується біспінорне представлення групи Лоренца. Докладно це виділення розглянуте в \cite{Chudak:2016, Volkotrub:2015laa,Ptashynskiy2019MultiparticleFO}. Тензори з цього лінійного простору позначатимемо ${{\hat{\Psi }}_{{{s}_{1}},{{c}_{1}},{{c}_{2}},{{c}_{3}}}}.$ Оскільки будь-який адрон є безкольоровим, з лінійного простору ${{\hat{\Psi }}_{{{s}_{1}},{{c}_{1}},{{c}_{2}},{{c}_{3}}}}$ потрібно виділити інваріантний підпростір тензорів виду ${{\hat{\Psi }}_{{{s}_{1}},{{c}_{1}},{{c}_{2}},{{c}_{3}}}}={{\hat{\Psi }}_{{{s}_{1}}}}{{\varepsilon }_{{{c}_{1}},{{c}_{2}},{{c}_{3}}}},$ які перетворюються за тривіальним представленням групи $S{{U}_{c}}\left( 3 \right)$ (тут ${{\varepsilon }_{{{c}_{1}},{{c}_{2}},{{c}_{3}}}}$ позначено тривимірний символ Леві-Чивіти). Однак, таке виділення зручно провести пізніше, коли буде врахована взаємодія між кварками. 

Отже, розглянемо тричастинкову операторнозначну польову функцію ${{\hat{\Psi }}_{{{s}_{1}},{{c}_{1}},{{c}_{2}},{{c}_{3}}}}\left( q \right),$ де $q-$довільний стовпець \eqref{Stovpec_na_pidmnojini_odnochasnosti} на підмножині одночасності. В якості дії для такого поля можемо взяти величину: 
\begin{equation}\label{dija}
S=\int{{{d}^{10}}q}\left( {{g}^{{{a}_{1}}{{a}_{2}}}}\frac{\partial {{{\bar{\Psi }}}_{{{s}_{1}},{{c}_{1}},{{c}_{2}},{{c}_{3}}}}\left( q \right)}{\partial {{x}^{{{a}_{1}}}}}\frac{\partial {{\Psi }_{{{s}_{1}},{{c}_{1}},{{c}_{2}},{{c}_{3}}}}\left( q \right)}{\partial {{x}^{{{a}_{2}}}}}-{{\left( 3{{m}_{q}} \right)}^{2}}{{{\bar{\Psi }}}_{{{s}_{1}},{{c}_{1}},{{c}_{2}},{{c}_{3}}}}\left( q \right){{\Psi }_{{{s}_{1}},{{c}_{1}},{{c}_{2}},{{c}_{3}}}}\left( q \right) \right).
\end{equation}
Тут ${{\bar{\Psi }}_{{{s}_{1}},{{c}_{1}},{{c}_{2}},{{c}_{3}}}}\left( q \right)$ позначає дираківські спряжене поле. По індексам, що повторюються, мається на увазі підсумування. Через $ {{m}_{q}} $ позначена маса констітуєнтного кварка, яку внаслідок знехтування всіма взаємодіями ми приймаємо однаковою для кварків всіх ароматів.

\begin{equation}\label{Lagrang_Eiler_dla_S0}
-{{\left( {{g}^{Minc}} \right)}^{ab}}\frac{{{\partial }^{2}}{{{\hat{\Psi }}}_{{{s}_{1}},{{c}_{1}}{{c}_{2}}{{c}_{3}}}}\left( X,{{{\vec{y}}}_{1}},{{{\vec{y}}}_{2}} \right)}{\partial {{X}^{a}}\partial {{X}^{b}}}-\left( {{\left( 3{{m}_{q}} \right)}^{2}}-\frac{9}{2}{{\Delta }_{{{{\vec{y}}}_{1}}}}-6{{\Delta }_{{{{\vec{y}}}_{2}}}} \right){{\hat{\Psi }}_{{{s}_{1}},{{c}_{1}}{{c}_{2}}{{c}_{3}}}}\left( X,{{{\vec{y}}}_{1}},{{{\vec{y}}}_{2}} \right)=0.
\end{equation}
Тут використано позначення: 
\begin{equation}\label{Poznachenna_Laplas}
{{\Delta }_{{{{\vec{y}}}_{a}}}}=\frac{{{\partial }^{2}}}{\partial {{\left( y_{a}^{1} \right)}^{2}}}+\frac{{{\partial }^{2}}}{\partial {{\left( y_{a}^{2} \right)}^{2}}}+\frac{{{\partial }^{2}}}{\partial {{\left( y_{a}^{3} \right)}^{2}}},a=1,2.
\end{equation}
Рівняння \eqref{Lagrang_Eiler_dla_S0} зручно переписати в виді:
\begin{equation}\label{Lagrang_Eiler_s_videleniem_vnutihnogo_hamiltonianu}
\begin{split}
   & -{{\left( {{g}^{Minc}} \right)}^{ab}}\frac{{{\partial }^{2}}{{{\hat{\Psi }}}_{{{s}_{1}},{{c}_{1}}{{c}_{2}}{{c}_{3}}}}\left( X,{{{\vec{y}}}_{1}},{{{\vec{y}}}_{2}} \right)}{\partial {{X}^{a}}\partial {{X}^{b}}}- \\ 
 & -\left( {{\left( 3{{m}_{q}} \right)}^{2}}+2\left( 3{{m}_{q}} \right)\left( -\frac{1}{2\left( \left( {2}/{3}\; \right){{m}_{q}} \right)}{{\Delta }_{{{{\vec{y}}}_{1}}}}-\frac{1}{2\left( \left( {1}/{2}\; \right){{m}_{q}} \right)}{{\Delta }_{{{{\vec{y}}}_{2}}}} \right) \right){{{\hat{\Psi }}}_{{{s}_{1}},{{c}_{1}}{{c}_{2}}{{c}_{3}}}}\left( X,{{{\vec{y}}}_{1}},{{{\vec{y}}}_{2}} \right)=0. \\ 
\end{split}
\end{equation}
Вираз $ \left( -\frac{1}{2\left( \left( {2}/{3}\; \right){{m}_{q}} \right)}{{\Delta }_{{{{\vec{y}}}_{1}}}}-\frac{1}{2\left( \left( {1}/{2}\; \right){{m}_{q}} \right)}{{\Delta }_{{{{\vec{y}}}_{2}}}} \right) $ співпадає із внутрішнім гамільтоніаном системи трьох невзаємодіючих нерелятивістських частинок. Оскільки, розглядається система фіксованої кількості частинок, залежність поля $ {{\hat{\Psi }}_{{{s}_{1}},{{c}_{1}}{{c}_{2}}{{c}_{3}}}}\left( X,{{{\vec{y}}}_{1}},{{{\vec{y}}}_{2}} \right) $ від внутрішніх змінних ${{\vec{y}}_{1}},{{\vec{y}}_{2}}$ повинна мати ненульові компоненти розкладу лише по тих власних функціях цього внутрішнього гамільтоніану, яким відповідають власні значення менші за ${m}_{q}.$ Тоді, з точністю до квадрату відношення власних значень внутрішнього гамільтоніану до $3{m}_{q}$ можемо записати рівняння \eqref{Lagrang_Eiler_dla_S0} в виді:
\begin{equation}\label{KGF}
-{{\left( {{g}^{Minc}} \right)}^{ab}}\frac{{{\partial }^{2}}{{{\hat{\Psi }}}_{{{s}_{1}},{{c}_{1}}{{c}_{2}}{{c}_{3}}}}\left( X,{{{\vec{y}}}_{1}},{{{\vec{y}}}_{2}} \right)}{\partial {{X}^{a}}\partial {{X}^{b}}}-{{\left( {{{\hat{H}}}^{\text{internal,0}}}\left( {{{\vec{y}}}_{1}},{{{\vec{y}}}_{2}} \right) \right)}^{2}}{{\hat{\Psi }}_{{{s}_{1}},{{c}_{1}}{{c}_{2}}{{c}_{3}}}}\left( X,{{{\vec{y}}}_{1}},{{{\vec{y}}}_{2}} \right)=0.
\end{equation}
Тут введене позначення:
\begin{equation}\label{H_internal}
{{\hat{H}}^{\text{internal,0}}}\left( {{{\vec{y}}}_{1}},{{{\vec{y}}}_{2}} \right)=3{{m}_{q}}\hat{E}-\frac{1}{2\left( \left( {2}/{3}\; \right){{m}_{q}} \right)}{{\Delta }_{{{{\vec{y}}}_{1}}}}-\frac{1}{2\left( \left( {1}/{2}\; \right){{m}_{q}} \right)}{{\Delta }_{{{{\vec{y}}}_{2}}}},
\end{equation}
де $\hat{E}-$одиничний оператор.
Оскільки, оператори ${{\hat{P}}_{a}}=i\left( {\partial }/{\partial {{X}^{a}}}\; \right)$ і ${{\hat{H}}^{\text{internal}}}\left( {{{\vec{y}}}_{1}},{{{\vec{y}}}_{2}} \right)$ комутують між собою, то для рівняння \eqref{KGF} може бути реалізована процедура <<розкладання на множники>> \cite{Bogolyubov_rus}, яка призводить до рівняння:
\begin{equation}\label{Trichastincove_rivnanna_Diraca}
i\hat{\gamma }_{{{s}_{1}}{{s}_{2}}}^{a}\frac{\partial {{{\hat{\Psi }}}_{{{s}_{2}},{{c}_{1}},{{c}_{2}},{{c}_{3}}}}\left( X,{{{\vec{y}}}_{1}},{{{\vec{y}}}_{2}} \right)}{\partial {{X}^{a}}}-{{\hat{H}}^{\text{internal,0}}}\left( {{{\vec{y}}}_{1}},{{{\vec{y}}}_{2}} \right){{\hat{\Psi }}_{{{s}_{1}},{{c}_{1}},{{c}_{2}},{{c}_{3}}}}\left( X,{{{\vec{y}}}_{1}},{{{\vec{y}}}_{2}} \right)=0,
\end{equation}
яке природно назвати тричастинковим рівнянням Дірака. Тут $ \hat{\gamma }_{{{s}_{1}}{{s}_{2}}}^{a}- $ елементи матриць Дірака. Рівняння \eqref{Trichastincove_rivnanna_Diraca} породжується лагранжіаном:
\begin{equation}\label{Lagranjian_Diraca_3Particls}
\begin{split}
    & L^{\left( 0\right)  } =\frac{i}{2}\left( \sum\limits_{b=0}^{3}{\left( {{{\hat{\bar{\Psi }}}}_{{{s}_{1}},{{c}_{1}}{{c}_{2}}{{c}_{3}}}}\left( q \right)\gamma _{{{s}_{1}}{{s}_{2}}}^{b}\frac{\partial {{{\hat{\Psi }}}_{{{s}_{2}},{{c}_{1}}{{c}_{2}}{{c}_{3}}}}\left( q \right)}{\partial {{q}^{b}}}-\frac{\partial {{{\hat{\bar{\Psi }}}}_{{{s}_{1}},{{c}_{1}}{{c}_{2}}{{c}_{3}}}}\left( q \right)}{\partial {{q}^{b}}}\gamma _{{{s}_{1}}{{s}_{2}}}^{b}{{{\hat{\Psi }}}_{{{s}_{2}},{{c}_{1}}{{c}_{2}}{{c}_{3}}}}\left( q \right) \right)} \right)- \\ 
  &-\left( 3{{m}_{q}} \right){{{\hat{\bar{\Psi }}}}_{{{s}_{1}},{{c}_{1}}{{c}_{2}}{{c}_{3}}}}\left( q \right){{{\hat{\Psi }}}_{{{s}_{1}},{{c}_{1}}{{c}_{2}}{{c}_{3}}}}\left( q \right)+\frac{1}{2\left( 3{{m}_{q}} \right)}\sum\limits_{b=4}^{9}{\sum\limits_{d=4}^{9}{{{g}^{bd}}\frac{\partial {{{\hat{\bar{\Psi }}}}_{{{s}_{1}},{{c}_{1}}{{c}_{2}}{{c}_{3}}}}\left( q \right)}{\partial {{q}^{b}}}\frac{\partial {{{\hat{\Psi }}}_{{{s}_{1}},{{c}_{1}}{{c}_{2}}{{c}_{3}}}}\left( q \right)}{\partial {{q}^{d}}}}} .\\ 
\end{split}
\end{equation}

\section{Калібрувальні поля на підмножині одночасності}
Тепер ми можемо врахувати сильну взаємодію між кварками звичайним способом - переходячи від глобально $S{{U}_{c}}\left( 3 \right)$ симетричного виразу \eqref{Lagranjian_Diraca_3Particls} до відповідного локально симетричного лагранжіану шляхом заміни похідних на оператори коваріантного диференціювання:
\begin{equation}\label{Kovariantni_pohidni_sz_3_polami}
\begin{split}
  & {{{\hat{D}}}_{b}}\left( {{{\hat{\Psi }}}_{{{s}_{1}},{{c}_{1}}{{c}_{2}}{{c}_{3}}}}\left( q \right) \right)=\frac{\partial {{{\hat{\Psi }}}_{{{s}_{1}},{{c}_{1}}{{c}_{2}}{{c}_{3}}}}\left( q \right)}{\partial {{q}^{b}}}- \\ 
& -ig\hat{A}_{b,{{g}_{1}}}^{\left( 1 \right)}\left( q \right)\lambda _{{{c}_{1}}{{c}_{4}}}^{{{g}_{1}}}{{{\hat{\Psi }}}_{{{s}_{1}},{{c}_{4}}{{c}_{2}}{{c}_{3}}}}\left( q \right)-ig\hat{A}_{b,{{g}_{2}}}^{\left( 2 \right)}\left( q \right)\lambda _{{{c}_{2}}{{c}_{4}}}^{{{g}_{2}}}{{{\hat{\Psi }}}_{{{s}_{1}},{{c}_{1}}{{c}_{4}}{{c}_{3}}}}\left( q \right)-ig\hat{A}_{b,{{g}_{3}}}^{\left( 3 \right)}\left( q \right)\lambda _{{{c}_{3}}{{c}_{4}}}^{{{g}_{3}}}{{{\hat{\Psi }}}_{{{s}_{1}},{{c}_{1}}{{c}_{2}}{{c}_{4}}}}\left( q \right), \\ 
& \overline{{{D}_{b}}{{\Psi }_{{{s}_{1}},{{c}_{1}}{{c}_{2}}{{c}_{3}}}}\left( q \right)}=\frac{\partial {{{\hat{\bar{\Psi }}}}_{{{s}_{1}},{{c}_{1}}{{c}_{2}}{{c}_{3}}}}\left( q \right)}{\partial {{q}^{b}}}+ \\ 
& +ig\hat{A}_{b,{{g}_{1}}}^{\left( 1 \right)}\left( q \right){{{\hat{\bar{\Psi }}}}_{{{s}_{1}},{{c}_{4}}{{c}_{2}}{{c}_{3}}}}\left( q \right)\lambda _{{{c}_{4}}{{c}_{1}}}^{{{g}_{1}}}+ig\hat{A}_{b,{{g}_{2}}}^{\left( 2 \right)}\left( q \right){{{\hat{\bar{\Psi }}}}_{{{s}_{1}},{{c}_{1}}{{c}_{4}}{{c}_{3}}}}\left( q \right)\lambda _{{{c}_{4}}{{c}_{2}}}^{{{g}_{2}}}+ig\hat{A}_{b,{{g}_{3}}}^{\left( 3 \right)}\left( q \right){{{\hat{\bar{\Psi }}}}_{{{s}_{1}},{{c}_{1}}{{c}_{2}}{{c}_{4}}}}\left( q \right)\lambda _{{{c}_{4}}{{c}_{3}}}^{{{g}_{3}}}. \\ 
\end{split}
\end{equation}
Тут $g-$константа сильної взаємодії, $\lambda _{{{c}_{1}}{{c}_{2}}}^{{{g}_{1}}},{{g}_{1}}=1,2\ldots ,8,{{c}_{1}},{{c}_{2}}=1,2,3-$ елементи матриць Гелл-Манна, $ \hat{A}_{b,{{g}_{1}}}^{\left( 1 \right)}\left( q \right),\hat{A}_{b,{{g}_{1}}}^{\left( 2 \right)}\left( q \right),\hat{A}_{b,{{g}_{1}}}^{\left( 3 \right)}\left( q \right)- $оператори калібрувальних полів. Оскільки, простір внутрішніх індексів глюонних полів є евклідовим, то немає різниці між верхніми і нижніми індексами і ми пишемо внутрішні індекси матриць Гелл-Манна верхніми, а внутрішні індекси глюонних полів нижніми, суто для зручності позначень. Кожне з цих полів перетворюється при локальному $S{{U}_{c}}\left( 3 \right)-$ перетворенні за звичайним законом, який дозволяє компенсувати доданки, що виникають за рахунок залежності параметрів перетворення від координат. Звернемо увагу на наявність саме трьох калібрувальних полів, а не одного, як є в одночастинковій квантовій теорії поля. Умова локальної інваріантності лагранжіану визначає лише закон перетворення калібрувального поля при відповідному локальному перетворенні ферміонних полів. Внаслідок цього, визначається лагранжіан і динамічні рівняння для калібрувального поля. Три введених калібрувальних поля повинні мати один і той самий закон перетворення і динамічні рівняння. Але до цих однакових рівнянь можуть бути поставлені різні граничні умови, що призведе до того, що ми матимемо три різних розв'язки цих рівнянь. Оскільки, ми збираємось описувати зв'язаний стан трьох кварків, то нам доведеться суттєво враховувати граничні умови по внутрішнім змінним. 

Окрім того, тепер можемо врахувати що баріон, процеси народження і знищення якого ми розраховуємо описати тричастинковим біспінорним полем, повинен бути безкольоровим. Тому, з лінійного простору триіндексних тензорів $ {{\hat{\Psi }}_{{{s}_{1}},{{c}_{1}}{{c}_{2}}{{c}_{3}}}}\left( q \right) $ виділимо інваріантний підпростір, на якому реалізується тривіальне представлення групи $S{{U}_{c}}\left( 3 \right)$. Для цього покладемо:
\begin{equation}\label{Bezcolorovij_pidprostir}
{{\hat{\Psi }}_{{{s}_{1}},{{c}_{1}}{{c}_{2}}{{c}_{3}}}}\left( q \right)={{\hat{\Psi }}_{{{s}_{1}}}}\left( q \right){{\varepsilon }_{{{c}_{1}}{{c}_{2}}{{c}_{3}}}}.
\end{equation}   
Це означає, що проєкції поля $ {{\hat{\Psi }}_{{{s}_{1}},{{c}_{1}}{{c}_{2}}{{c}_{3}}}}\left( q \right) $ на решту інваріантних підпросторів, з яких складається лінійний простір триіндексних тензорів ми поклали рівними нулю, виражаючи тим самим відсутність кольору у баріону. 

Замінюючи в \eqref{Lagranjian_Diraca_3Particls} похідні на оператори \eqref{Kovariantni_pohidni_sz_3_polami} і враховуючи \eqref{Bezcolorovij_pidprostir} отримаємо:
\begin{equation}\label{Lagrangian_z_podovjenimi_pohidnimi}
\begin{split}
  & \frac{L}{6}=\frac{i}{2}\left( \sum\limits_{b=0}^{3}{\left( {{{\hat{\bar{\Psi }}}}_{{{s}_{1}}}}\left( q \right)\gamma _{{{s}_{1}}{{s}_{2}}}^{b}\frac{\partial {{{\hat{\Psi }}}_{{{s}_{2}}}}\left( q \right)}{\partial {{q}^{b}}}-\frac{\partial {{{\hat{\bar{\Psi }}}}_{{{s}_{1}}}}\left( q \right)}{\partial {{q}^{b}}}\gamma _{{{s}_{1}}{{s}_{2}}}^{b}{{{\hat{\Psi }}}_{{{s}_{2}}}}\left( q \right) \right)} \right)- \\ 
& -\left( 3{{m}_{q}} \right){{{\hat{\bar{\Psi }}}}_{{{s}_{1}}}}\left( q \right){{{\hat{\Psi }}}_{{{s}_{1}}}}\left( q \right)+\frac{1}{2\left( 3{{m}_{q}} \right)}\sum\limits_{b=4}^{9}{\sum\limits_{d=4}^{9}{{{g}^{bd}}\frac{\partial {{{\hat{\bar{\Psi }}}}_{{{s}_{1}}}}\left( q \right)}{\partial {{q}^{b}}}\frac{\partial {{{\hat{\Psi }}}_{{{s}_{1}}}}\left( q \right)}{\partial {{q}^{d}}}}}+ \\ 
&+\frac{{{g}^{2}}}{\left( 9{{m}_{q}} \right)}{{{\hat{\bar{\Psi }}}}_{{{s}_{1}}}}\left( q \right){{{\hat{\Psi }}}_{{{s}_{1}}}}\left( q \right)\times  \\ 
& \times \sum\limits_{b=4}^{9}{\sum\limits_{d=4}^{9}{{{g}^{bd}}}}\left( \hat{A}_{b,{{g}_{1}}}^{\left( 1 \right)}\left( q \right)\hat{A}_{d,{{g}_{1}}}^{\left( 1 \right)}\left( q \right)+\hat{A}_{b,{{g}_{1}}}^{\left( 2 \right)}\left( q \right)\hat{A}_{d,{{g}_{1}}}^{\left( 2 \right)}\left( q \right)+\hat{A}_{b,{{g}_{1}}}^{\left( 3 \right)}\left( q \right)\hat{A}_{d,{{g}_{1}}}^{\left( 3 \right)}\left( q \right)- \right. \\ 
& \left. -\hat{A}_{b,{{g}_{1}}}^{\left( 1 \right)}\left( q \right)\hat{A}_{d,{{g}_{1}}}^{\left( 2 \right)}\left( q \right)-\hat{A}_{b,{{g}_{1}}}^{\left( 1 \right)}\left( q \right)\hat{A}_{d,{{g}_{1}}}^{\left( 3 \right)}\left( q \right)-\hat{A}_{b,{{g}_{1}}}^{\left( 2 \right)}\left( q \right)\hat{A}_{d,{{g}_{1}}}^{\left( 3 \right)}\left( q \right) \right) \\ 
\end{split}
\end{equation}
Тут ми для зручності розділили лагранжіан на множник 6, який виникає за рахунок підсумування компонент символу Леві-Чивіти. 

Введемо тепер замість $\hat A_{b,{{g}_{1}}}^{\left( 1 \right)}\left( q \right),\hat A_{b,{{g}_{1}}}^{\left( 2 \right)}\left( q \right),\hat A_{b,{{g}_{1}}}^{\left( 3 \right)}\left( q \right) $ нові калібрувальні поля $ \hat A_{b,{{g}_{1}}}^{\left( + \right)}\left( q \right), \hat A_{b,{{g}_{1}}}^{\left( -,1 \right)}\left( q \right),\hat A_{b,{{g}_{1}}}^{\left( -,2 \right)}\left( q \right) $ за допомогою співвідношень, аналогічних тим, які визначають тричастинкові координати Якобі:
\begin{equation}\label{Pola_Aplus_i_Aminus1_2}
\begin{split}
 &\hat A_{b,{{g}_{1}}}^{\left( + \right)}\left( q \right)=\frac{1}{3}\left(\hat A_{b,{{g}_{1}}}^{\left( 1 \right)}\left( q \right)+\hat A_{b,{{g}_{1}}}^{\left( 2 \right)}\left( q \right)+\hat A_{b,{{g}_{1}}}^{\left( 3 \right)}\left( q \right) \right) \\ 
&\hat A_{b,{{g}_{1}}}^{\left( -,1 \right)}\left( q \right)=\hat A_{b,{{g}_{1}}}^{\left( 3 \right)}\left( q \right)-\frac{1}{2}\left( \hat A_{b,{{g}_{1}}}^{\left( 1 \right)}\left( q \right)+\hat A_{b,{{g}_{1}}}^{\left( 2 \right)}\left( q \right) \right) \\ 
& \hat A_{b,{{g}_{1}}}^{\left( -,2 \right)}\left( q \right)=\hat A_{b,{{g}_{1}}}^{\left( 2 \right)}\left( q \right)-\hat A_{b,{{g}_{1}}}^{\left( 1 \right)}\left( q \right) \\ 
\end{split}
\end{equation}
Обернені співвідношення мають вид:
\begin{equation}\label{Oberneni_spivvidnochenna}
\begin{split}
  &\hat A_{b,{{g}_{1}}}^{\left( 1 \right)}\left( q \right)=\hat A_{b,{{g}_{1}}}^{\left( + \right)}\left( q \right)-\frac{1}{3}\hat A_{b,{{g}_{1}}}^{\left( -,1 \right)}\left( q \right)-\frac{1}{2}\hat A_{b,{{g}_{1}}}^{\left( -,2 \right)}\left( q \right) \\ 
& \hat A_{b,{{g}_{1}}}^{\left( 2 \right)}\left( q \right)=\hat A_{b,{{g}_{1}}}^{\left( + \right)}\left( q \right)-\frac{1}{3} \hat A_{b,{{g}_{1}}}^{\left( -,1 \right)}\left( q \right)+\frac{1}{2}\hat A_{b,{{g}_{1}}}^{\left( -,2 \right)}\left( q \right) \\ 
&\hat A_{b,{{g}_{1}}}^{\left( 3 \right)}\left( q \right)=\hat A_{b,{{g}_{1}}}^{\left( + \right)}\left( q \right)+\frac{2}{3}\hat A_{b,{{g}_{1}}}^{\left( -,1 \right)}\left( q \right) \\ 
\end{split}
\end{equation}
Після заміни \eqref{Oberneni_spivvidnochenna} лагранжіан \eqref{Lagrangian_z_podovjenimi_pohidnimi} може бути записаний у виді:
\begin{equation}\label{Lagranjian_pisla_zamini}
\begin{split}
  & \frac{L}{6}=\frac{i}{2}\left( \sum\limits_{b=0}^{3}{\left( {{{\hat{\bar{\Psi }}}}_{{{s}_{1}}}}\left( q \right)\gamma _{{{s}_{1}}{{s}_{2}}}^{b}\frac{\partial {{{\hat{\Psi }}}_{{{s}_{2}}}}\left( q \right)}{\partial {{q}^{b}}}-\frac{\partial {{{\hat{\bar{\Psi }}}}_{{{s}_{1}}}}\left( q \right)}{\partial {{q}^{b}}}\gamma _{{{s}_{1}}{{s}_{2}}}^{b}{{{\hat{\Psi }}}_{{{s}_{2}}}}\left( q \right) \right)} \right)- \\ 
& -\left( 3{{m}_{q}} \right){{{\hat{\bar{\Psi }}}}_{{{s}_{1}}}}\left( q \right){{{\hat{\Psi }}}_{{{s}_{1}}}}\left( q \right)+\frac{1}{2\left( 3{{m}_{q}} \right)}\sum\limits_{b=4}^{9}{\sum\limits_{d=4}^{9}{{{g}^{bd}}\frac{\partial {{{\hat{\bar{\Psi }}}}_{{{s}_{1}}}}\left( q \right)}{\partial {{q}^{b}}}\frac{\partial {{{\hat{\Psi }}}_{{{s}_{1}}}}\left( q \right)}{\partial {{q}^{d}}}}}+ \\ 
& +\frac{{{g}^{2}}}{\left( 9{{m}_{q}} \right)}\left( \sum\limits_{b=4}^{9}{\sum\limits_{d=4}^{9}{{{g}^{bd}}}}\left( \hat{A}_{b,{{g}_{1}}}^{\left( -,1 \right)}\left( q \right)\hat{A}_{d,{{g}_{1}}}^{\left( -,1 \right)}\left( q \right)+\frac{3}{4}\hat{A}_{b,{{g}_{1}}}^{\left( -,2 \right)}\left( q \right)\hat{A}_{d,{{g}_{1}}}^{\left( -,2 \right)}\left( q \right) \right) \right){{{\hat{\bar{\Psi }}}}_{{{s}_{1}}}}\left( q \right){{{\hat{\Psi }}}_{{{s}_{1}}}}\left( q \right) \\ 
\end{split}
\end{equation}

\section{Узагальнення способу досягнення калібрувальної інваріантності}
Зазначимо, що використаний спосіб отримання лагранжіану, інваріантного відносно локального $S{{U}_{c}}\left( 3 \right)$ перетворення не є найбільш загальним. Дійсно, якщо ми розглянемо вираз, який отримується з $\frac{\partial {{{\hat{\bar{\Psi }}}}_{{{s}_{1}},{{c}_{1}}{{c}_{2}}{{c}_{3}}}}\left( q \right)}{\partial {{q}^{d}}}\frac{\partial {{{\hat{\Psi }}}_{{{s}_{1}},{{c}_{1}}{{c}_{2}}{{c}_{3}}}}\left( q \right)}{\partial {{q}^{b}}}$ шляхом заміни похідних на оператори коваріантного диференціювання, то ми матимемо доданки трьох типів: які не містять матриць Гелл-Манна, які містять матричні елементи однієї матриці $\lambda _{{{c}_{1}}{{c}_{2}}}^{{{g}_{1}}}$ і які містять добутки матричних елементів двох матриць $\lambda _{{{c}_{1}}{{c}_{2}}}^{{{g}_{1}}}\lambda _{{{c}_{3}}{{c}_{4}}}^{{{g}_{2}}}.$ Перші ступені матричних елементів матриць Гелл-Манна входять в лагранжіан у виді згорток із калібрувальними полями $\hat{A}_{{{g}_{1}}}^{\left( n \right)}\left( q \right)\lambda _{{{c}_{1}}{{c}_{2}}}^{{{g}_{1}}},n=1,2,3$, а другі - у виді згорток $\hat{A}_{{{g}_{1}},b}^{\left( {{n}_{1}} \right)}\left( q \right)\hat{A}_{{{g}_{2}},d}^{\left( {{n}_{2}} \right)}\left( q \right)\lambda _{{{c}_{1}}{{c}_{2}}}^{{{g}_{1}}}\lambda _{{{c}_{3}}{{c}_{4}}}^{{{g}_{2}}}.$ При цьому, для досягнення інваріантності відносно локального $S{{U}_{c}}\left( 3 \right)-$перетворення суттєвий не конкретний вираз цих коефіцієнтів, а закон їх перетворення. Тому, якщо ми добуток  $\hat{A}_{{{g}_{1}},b}^{\left( {{n}_{1}} \right)}\left( q \right)\hat{A}_{{{g}_{2}},d}^{\left( {{n}_{2}} \right)}\left( q \right)$ замінимо на тензор $\hat{A}_{bd,{{g}_{1}}{{g}_{2}}}^{\left( {{n}_{1}},{{n}_{2}} \right)}\left( q \right),$ який має той самий закон перетворення що й добуток $\hat{A}_{{{g}_{1}},b}^{\left( {{n}_{1}} \right)}\left( q \right)\hat{A}_{{{g}_{2}},d}^{\left( {{n}_{2}} \right)}\left( q \right),$ то отримаємо більш загальний вираз лагранжіану, який задовольняє вимозі локальної $S{{U}_{c}}\left( 3 \right)-$ інваріантності. 

При локальному $S{{U}_{c}}\left( 3 \right)-$перетворенні: 
\begin{equation}\label{Localne_peretvorenna_SU3c}
\begin{split}
  & {{{{\hat{\Psi }}'}}_{{{s}_{1}},{{c}_{1}}{{c}_{2}}{{c}_{3}}}}\left( q \right)={{{\hat{U}}}_{{{c}_{1}}{{c}_{4}}}}\left( q \right){{{\hat{U}}}_{{{c}_{2}}{{c}_{5}}}}\left( q \right){{{\hat{U}}}_{{{c}_{3}}{{c}_{6}}}}\left( q \right){{{\hat{\Psi }}}_{{{s}_{1}},{{c}_{4}}{{c}_{5}}{{c}_{6}}}}\left( q \right), \\ 
& {{{{\hat{\bar{\Psi }}}'}}_{{{s}_{1}},{{c}_{1}}{{c}_{2}}{{c}_{3}}}}\left( q \right)={{{\hat{\bar{\Psi }}}}_{{{s}_{1}},{{c}_{4}}{{c}_{5}}{{c}_{6}}}}\left( q \right)\hat{U}_{{{c}_{4}}{{c}_{1}}}^{-1}\left( q \right)\hat{U}_{{{c}_{5}}{{c}_{2}}}^{-1}\left( q \right)\hat{U}_{{{c}_{6}}{{c}_{3}}}^{-1}\left( q \right), \\ 
\end{split}
\end{equation}
де $ \hat{U}\left( q \right)=\exp \left( i{{{\hat{\lambda }}}^{{{g}_{1}}}}{{\theta }_{{{g}_{1}}}}\left( q \right) \right), $ і 
${{\theta }_{{{g}_{1}}}}\left( q \right),-$залежні від координат параметри перетворення, ($ {{g}_{1}}=1,2,\ldots ,8$) калібрувальні поля перетворюються за законом 
\begin{equation}\label{Zacon_odnogluonnnogo_pola}
\hat{{A}'}_{b,{{g}_{1}}}^{\left( n \right)}\left( q \right)={{D}_{{{g}_{1}}{{g}_{2}}}}\left( q \right)\hat{A}_{b,{{g}_{2}}}^{\left( n \right)}\left( q \right)+\frac{\partial {{\theta }_{{{g}_{1}}}}\left( q \right)}{\partial {{q}^{b}}}. 
\end{equation}
Тут ${{D}_{{{g}_{1}}{{g}_{2}}}}\left( q \right)-$ елементи матриць приєднаного представлення групи $S{{U}_{c}}\left( 3 \right).$  Відповідно, добуток двох компонент калібрувального поля перетворюється за законом:
\begin{equation}\label{Peretvorenna_dobutcu}
\begin{split}
& \hat{{A}'}_{b,{{g}_{1}}}^{\left( {{n}_{1}} \right)}\left( q \right)\hat{{A}'}_{d,{{g}_{3}}}^{\left( {{n}_{2}} \right)}\left( q \right)={{D}_{{{g}_{1}}{{g}_{2}}}}\left( q \right){{D}_{{{g}_{3}}{{g}_{4}}}}\left( q \right)\hat{A}_{b,{{g}_{2}}}^{\left( {{n}_{1}} \right)}\left( q \right)\hat{A}_{d,{{g}_{4}}}^{\left( {{n}_{2}} \right)}\left( q \right)+ \\ 
& +{{D}_{{{g}_{1}}{{g}_{2}}}}\left( q \right)\hat{A}_{b,{{g}_{2}}}^{\left( {{n}_{1}} \right)}\left( q \right)\frac{\partial {{\theta }_{{{g}_{3}}}}\left( q \right)}{\partial {{q}^{d}}}+\frac{\partial {{\theta }_{{{g}_{1}}}}\left( q \right)}{\partial {{q}^{b}}}{{D}_{{{g}_{3}}{{g}_{4}}}}\left( q \right)\hat{A}_{d,{{g}_{4}}}^{\left( {{n}_{2}} \right)}\left( q \right)+\frac{\partial {{\theta }_{{{g}_{1}}}}\left( q \right)}{\partial {{q}^{b}}}\frac{\partial {{\theta }_{{{g}_{3}}}}\left( q \right)}{\partial {{q}^{d}}} \\ 
\end{split}
\end{equation}
Тому, якщо для тензору $ \hat{A}_{bd,{{g}_{1}}{{g}_{3}}}^{\left( {{n}_{1}},{{n}_{2}} \right)}\left( q \right) $ буде виконуватися закон перетворення:
\begin{equation}\label{Zacon_peretvorenna_dla_dvogluonnogo_tenzora}
\begin{split}
  & \hat{{A}'}_{bd,{{g}_{1}}{{g}_{3}}}^{\left( {{n}_{1}}{{n}_{2}} \right)}\left( q \right)={{D}_{{{g}_{1}}{{g}_{2}}}}\left( q \right){{D}_{{{g}_{3}}{{g}_{4}}}}\left( q \right)\hat{A}_{bd,{{g}_{2}}{{g}_{4}}}^{\left( {{n}_{1}}{{n}_{2}} \right)}\left( q \right)+ \\ 
& +{{D}_{{{g}_{1}}{{g}_{2}}}}\left( q \right)\hat{A}_{b,{{g}_{2}}}^{\left( {{n}_{1}} \right)}\left( q \right)\frac{\partial {{\theta }_{{{g}_{3}}}}\left( q \right)}{\partial {{q}^{d}}}+\frac{\partial {{\theta }_{{{g}_{1}}}}\left( q \right)}{\partial {{q}^{b}}}{{D}_{{{g}_{3}}{{g}_{4}}}}\left( q \right)\hat{A}_{d,{{g}_{4}}}^{\left( {{n}_{2}} \right)}\left( q \right)+\frac{\partial {{\theta }_{{{g}_{1}}}}\left( q \right)}{\partial {{q}^{b}}}\frac{\partial {{\theta }_{{{g}_{3}}}}\left( q \right)}{\partial {{q}^{d}}}, \\ 
\end{split}
\end{equation}
то лагранжіан, в якому добуток компонент калібрувальних полів замінений на тензор, буде інваріантним відносно локального $S{{U}_{c}}\left( 3 \right)$ перетворення. Поле, яке описується тензором $ \hat{A}_{bd,{{g}_{1}}{{g}_{3}}}^{\left( {{n}_{1}},{{n}_{2}} \right)}\left( q \right) $ будемо називати двоглюонним полем. Далі ми покажемо, спираючись на результати робіт \cite{Korotca_statta_v_UJP, Ptashynskiy2019MultiparticleFO}, що це поле описує процеси народження і знищення зв'язаного стану двох глюонів, що взаємодіють із трьома кварками. Такі зв'язані стани глюонів, як відомо, прийнято називати глюболами.

Як видно з закону перетворення \eqref{Zacon_peretvorenna_dla_dvogluonnogo_tenzora}, закон перетворення двоглюонного поля містить оператори одночастинкового глюонного поля. Проте, як видно з попередніх розрахунків для випадку, коли кольоровий стан трикваркової системи описується тензором Леві-Чивіти, доданки що містять перші ступені елементів матриць Гелл-Манна випадають із виразу для лагранжіану, внаслідок того, що згортка по кольоровим індексам призводить до сліду $Sp\left( {{{\hat{\lambda }}}^{{{g}_{1}}}} \right)$. Оскільки, одночастинкові поля входять у такому випадку в вираз для лагранжіану в виді $ \hat{A}_{{{g}_{1}}}^{\left( n \right)}\left( q \right)Sp\left( {{{\hat{\lambda }}}^{{{g}_{1}}}} \right),$ то вони теж випадають із лагранжіану. Тобто, в моделі, що розглядається для випадку безкольорового стану баріону такі поля стають нефізичними, бо оскільки вони не входять у лагранжіан, їх значення не визначаються динамікою системи. Окрім того, так як одночастинкові глюонні поля не входять у лагранжіан, для його інваріантності відносно локальних $S{{U}_{c}}\left( 3 \right)-$перетворень, зникає необхідність перетворювати ці поля за законом \eqref{Zacon_odnogluonnnogo_pola}. Тому, в усіх можливих калібруваннях ці поля можна покласти рівними нулю.
Тоді, замість \eqref{Zacon_peretvorenna_dla_dvogluonnogo_tenzora} отримаємо:
\begin{equation}\label{Zacon_dvogluonnogo_pola_bez_odnochastincovih_poliv}
\hat{{A}'}_{bd,{{g}_{1}}{{g}_{3}}}^{\left( {{n}_{1}}{{n}_{2}} \right)}\left( q \right)={{D}_{{{g}_{1}}{{g}_{2}}}}\left( q \right){{D}_{{{g}_{3}}{{g}_{4}}}}\left( q \right)\hat{A}_{bd,{{g}_{2}}{{g}_{4}}}^{\left( {{n}_{1}}{{n}_{2}} \right)}\left( q \right)+\frac{\partial {{\theta }_{{{g}_{1}}}}\left( q \right)}{\partial {{q}^{b}}}\frac{\partial {{\theta }_{{{g}_{3}}}}\left( q \right)}{\partial {{q}^{d}}}.
\end{equation}
При цьому лагранжіан, замість виразу \eqref{Lagrangian_z_podovjenimi_pohidnimi} прийме вид: 
\begin{equation}\label{Lagrangian_z_dvogluonnimi_polami}
\begin{split}
& \frac{L}{6}=\frac{i}{2}\left( \sum\limits_{b=0}^{3}{\left( {{{\hat{\bar{\Psi }}}}_{{{s}_{1}}}}\left( q \right)\gamma _{{{s}_{1}}{{s}_{2}}}^{b}\frac{\partial {{{\hat{\Psi }}}_{{{s}_{2}}}}\left( q \right)}{\partial {{q}^{b}}}-\frac{\partial {{{\hat{\bar{\Psi }}}}_{{{s}_{1}}}}\left( q \right)}{\partial {{q}^{b}}}\gamma _{{{s}_{1}}{{s}_{2}}}^{b}{{{\hat{\Psi }}}_{{{s}_{2}}}}\left( q \right) \right)} \right)- \\ 
& -\left( 3{{m}_{q}} \right){{{\hat{\bar{\Psi }}}}_{{{s}_{1}}}}\left( q \right){{{\hat{\Psi }}}_{{{s}_{1}}}}\left( q \right)-\frac{1}{2\left( 3{{m}_{q}} \right)}\sum\limits_{b=4}^{9}{\sum\limits_{d=4}^{9}{{{g}^{bd}}\frac{\partial {{{\hat{\bar{\Psi }}}}_{{{s}_{1}}}}\left( q \right)}{\partial {{q}^{b}}}\frac{\partial {{{\hat{\Psi }}}_{{{s}_{1}}}}\left( q \right)}{\partial {{q}^{d}}}}}- \\ 
& -\frac{{{g}^{2}}}{\left( 9{{m}_{q}} \right)}{{{\hat{\bar{\Psi }}}}_{{{s}_{1}}}}\left( q \right){{{\hat{\Psi }}}_{{{s}_{1}}}}\left( q \right)\times  \\ 
& \times \sum\limits_{b=4}^{9}{\sum\limits_{d=4}^{9}{{{g}^{bd}}}}{{\delta }^{{{g}_{1}}{{g}_{2}}}}\left( \hat{A}_{bd,{{g}_{1}}{{g}_{2}}}^{\left( 1,1 \right)}\left( q \right)+\hat{A}_{bd,{{g}_{1}}{{g}_{2}}}^{\left( 2,2 \right)}\left( q \right)+\hat{A}_{bd,{{g}_{1}}{{g}_{2}}}^{\left( 3,3 \right)}\left( q \right) - \right. \\ 
& \left. -\hat{A}_{bd,{{g}_{1}}{{g}_{2}}}^{\left( 1,2 \right)}\left( q \right)-\hat{A}_{bd,{{g}_{1}}{{g}_{2}}}^{\left( 1,3 \right)}\left( q \right)-\hat{A}_{bd,{{g}_{1}}{{g}_{2}}}^{\left( 2,3 \right)}\left( q \right) \right) \\ 
\end{split}
\end{equation}
Дельта-символ Кронекера ${{\delta }^{{{g}_{1}}{{g}_{2}}}}$ походить від сліду добутку двох матриць Гелл-Манна. Введемо далі позначення:
\begin{equation}\label{Poznachenna_Aplus_Aminus_V}
\begin{split}
  & \hat{A}_{bd,{{g}_{1}}{{g}_{2}}}^{\left( + \right)}\left( q \right)=\hat{A}_{bd,{{g}_{1}}{{g}_{2}}}^{\left( 1,1 \right)}\left( q \right)+\hat{A}_{bd,{{g}_{1}}{{g}_{2}}}^{\left( 2,2 \right)}\left( q \right)+\hat{A}_{bd,{{g}_{1}}{{g}_{2}}}^{\left( 3,3 \right)}\left( q \right), \\ 
& \hat{A}_{bd,{{g}_{1}}{{g}_{2}}}^{\left( - \right)}\left( q \right)=\hat{A}_{bd,{{g}_{1}}{{g}_{2}}}^{\left( 1,2 \right)}\left( q \right)+\hat{A}_{bd,{{g}_{1}}{{g}_{2}}}^{\left( 1,3 \right)}\left( q \right)+\hat{A}_{bd,{{g}_{1}}{{g}_{2}}}^{\left( 2,3 \right)}\left( q \right), \\ 
& {{{\hat{V}}}_{bd,{{g}_{1}}{{g}_{2}}}}\left( q \right)=\hat{A}_{bd,{{g}_{1}}{{g}_{2}}}^{\left( + \right)}\left( q \right)-\hat{A}_{bd,{{g}_{1}}{{g}_{2}}}^{\left( - \right)}\left( q \right), \\ 
& \hat{V}\left( q \right)=\sum\limits_{b=4}^{10}{\sum\limits_{d=4}^{10}{{{g}^{bd}}}{{\delta }^{{{g}_{1}}{{g}_{2}}}}}{{{\hat{V}}}_{bd,{{g}_{1}}{{g}_{2}}}}\left( q \right). \\
\end{split}
\end{equation}
Як видно з \eqref{Zacon_dvogluonnogo_pola_bez_odnochastincovih_poliv}, неоднорідні доданки в законі перетворення тензорних полів є однаковими для всіх типів цих полів. Тому, поле $ {{\hat{V}}_{bd,{{g}_{1}}{{g}_{2}}}}\left( q \right)$ перетворюється за тензорним добутком двох приєднаних представлень групи $S{{U}_{c}}\left( 3 \right)$, на відміну від одноглюонного поля, яке внаслідок наявності неоднорідного доданку не перетворюється ні за яким певним перетворенням цієї групи. Область значень поля $ \hat{V}\left( q \right)$ є проєкцією лінійного простору тензорів $ {{\hat{V}}_{bd,{{g}_{1}}{{g}_{2}}}} $ на інваріантний підпростір, на якому реалізується скалярне представлення групи перетворень при переході з однієї інерційної системи відліку до іншої, і скалярне представлення групи локальних $S{{U}_{c}}\left( 3 \right)$ перетворень. Група перетворень при зміні системи відліку, як це розглядалося в \cite{Ptashynskiy2019MultiparticleFO}, відрізняється від групи Лоренца тим, що на підпросторі внутрішніх змінних (у поля $ {{\hat{V}}_{bd,{{g}_{1}}{{g}_{2}}}}\left( q \right)$ і полів, з яких воно побудовано є ненульові компоненти лише на цьому підпросторі) бусти замінюються на тотожні перетворення. Лагранжіан \eqref{Lagrangian_z_dvogluonnimi_polami} приймає вид:
\begin{equation}\label{Lagrangian_z_polem_V}
\begin{split}
& \frac{L}{6}=\frac{i}{2}\left( \sum\limits_{b=0}^{3}{\left( {{{\hat{\bar{\Psi }}}}_{{{s}_{1}}}}\left( q \right)\gamma _{{{s}_{1}}{{s}_{2}}}^{b}\frac{\partial {{{\hat{\Psi }}}_{{{s}_{2}}}}\left( q \right)}{\partial {{q}^{b}}}-\frac{\partial {{{\hat{\bar{\Psi }}}}_{{{s}_{1}}}}\left( q \right)}{\partial {{q}^{b}}}\gamma _{{{s}_{1}}{{s}_{2}}}^{b}{{{\hat{\Psi }}}_{{{s}_{2}}}}\left( q \right) \right)} \right)- \\ 
& -\left( 3{{m}_{q}} \right){{{\hat{\bar{\Psi }}}}_{{{s}_{1}}}}\left( q \right){{{\hat{\Psi }}}_{{{s}_{1}}}}\left( q \right)+\frac{1}{2\left( 3{{m}_{q}} \right)}\sum\limits_{b=4}^{9}{\sum\limits_{d=4}^{9}{{{g}^{bd}}\frac{\partial {{{\hat{\bar{\Psi }}}}_{{{s}_{1}}}}\left( q \right)}{\partial {{q}^{b}}}\frac{\partial {{{\hat{\Psi }}}_{{{s}_{1}}}}\left( q \right)}{\partial {{q}^{d}}}}}- \\ 
& +\frac{{{g}^{2}}}{\left( 9{{m}_{q}} \right)}{{{\hat{\bar{\Psi }}}}_{{{s}_{1}}}}\left( q \right){{{\hat{\Psi }}}_{{{s}_{1}}}}\left( q \right)\hat{V}\left( q \right) \\ 
\end{split}
\end{equation}

\section{Динамічна модель взаємодіючих тричастинкового біспінорного поля і двоглюонного поля}
Лагранжіан \eqref{Lagrangian_z_polem_V}, вочевидь, потрібно доповнити лагранжіаном поля $ \hat{V}\left( q \right). $ Для цього розглянемо спочатку тензорне поле $ {{\hat{V}}_{bd,{{g}_{1}}{{g}_{2}}}}\left( q \right). $ Оскільки, це поле перетворюється за певним представленням локальної групи $S{{U}_{c}}\left( 3 \right)$, то воно аналогічне до <<полів матерії>> і до нього можна застосувати звичайний спосіб побудови лагранжіану. Лагранжіан вільного поля $ {{\hat{V}}_{bd,{{g}_{1}}{{g}_{2}}}}\left( q \right) $ може бути обраний у виді:
\begin{equation}\label{LV }
{{L}_{V}}=\frac{1}{2}{{g}^{b{{b}_{1}}}}{{g}^{d{{d}_{1}}}}{{g}^{l{{l}_{1}}}}\frac{\partial {{{\hat{V}}}_{bd,{{g}_{1}}{{g}_{2}}}}\left( q \right)}{\partial {{q}^{l}}}\frac{\partial {{{\hat{V}}}_{{{b}_{1}}{{d}_{1}},{{g}_{1}}{{g}_{2}}}}\left( q \right)}{\partial {{q}^{{{l}_{1}}}}}-\frac{1}{2}M_{G}^{2}{{g}^{b{{b}_{1}}}}{{g}^{d{{d}_{1}}}}{{\hat{V}}_{bd,{{g}_{1}}{{g}_{2}}}}\left( q \right){{\hat{V}}_{{{b}_{1}}{{d}_{1}},{{g}_{1}}{{g}_{2}}}}\left( q \right).
\end{equation}
Тут ${{M}_{G}}$ позначена маса частинок (глюболів), народження і знищення яких описують оператори поля $ {{\hat{V}}_{bd,{{g}_{1}}{{g}_{2}}}}\left( q \right). $ Цей лагранжіан не є інваріантним відносно локального перетворення полів згідно з тензорним добутком двох приєднаних представлень групи $S{{U}_{c}}\left( 3 \right).$ 
Такої інваріантності можна досягти використовуючи міркування, аналогічні тим, що призвели до лагранжіану \eqref{Lagrangian_z_dvogluonnimi_polami}. Спочатку замінимо в лагранжіані похідні на оператори коваріантного диференціювання, які для поля, що перетворюється за тензорним добутком двох приєднаних представлень групи $S{{U}_{c}}\left( 3 \right)$ мають вид:
\begin{equation}\label{Kovariantna_pohidna_v_priednanomu_predstavlenni}
\begin{split}
  & {{{\hat{D}}}_{l}}\left( {{{\hat{V}}}_{bd,{{g}_{1}}{{g}_{2}}}}\left( q \right) \right)=\frac{\partial {{{\hat{V}}}_{bd,{{g}_{1}}{{g}_{2}}}}\left( q \right)}{\partial {{q}^{l}}}- \\ 
& -ig\hat A_{l,{{g}_{3}}}^{\left( I \right)}\left( q \right)\hat{I}_{{{g}_{1}}{{g}_{4}}}^{{{g}_{3}}}{{{\hat{V}}}_{bd,{{g}_{4}}{{g}_{2}}}}\left( q \right)-ig \hat A_{l,{{g}_{5}}}^{\left( II \right)}\left( q \right)\hat{I}_{{{g}_{2}}{{g}_{6}}}^{{{g}_{5}}}{{{\hat{V}}}_{bd,{{g}_{1}}{{g}_{6}}}}\left( q \right). \\ 
\end{split}
\end{equation} 
Тут $\hat{I}_{{{g}_{2}}{{g}_{3}}}^{{{g}_{1}}}$- матричні елементи генераторів приєднаного представлення групи $S{{U}_{c}}\left( 3 \right)$, і  $\hat{A}_{l,{{g}_{1}}}^{\left( I \right)}\left( q \right), $ $\hat{A}_{l,{{g}_{1}}}^{\left( II \right)}\left( q \right)$- оператори калібрувальних полів, у якості яких можуть бути використані будь-які поля із законом перетворення \eqref{Zacon_odnogluonnnogo_pola}. В тому числі це можуть бути й одночастинкові калібрувальні поля $\hat{A}_{l,{{g}_{1}}}^{\left( n \right)}\left( q \right),n=1,2,3,$ розглянуті раніше, але тепер <<натягнуті>> не на генератори фундаментального представлення групи  $S{{U}_{c}}\left( 3 \right),$ а на генератори приєднаного представлення. Доданок лагранжіану, який містить згортку коваріантних похідних матиме вид:   
\begin{equation}\label{Zgortca_covariantnih_pohidnih}
\begin{split}
  & {{g}^{b{{b}_{1}}}}{{g}^{d{{d}_{1}}}}{{g}^{l{{l}_{1}}}}{{{\hat{D}}}_{l}}\left( {{{\hat{V}}}_{bd,{{g}_{1}}{{g}_{2}}}}\left( q \right) \right){{{\hat{D}}}_{{{l}_{1}}}}\left( {{{\hat{V}}}_{{{b}_{1}}{{d}_{1}},{{g}_{1}}{{g}_{2}}}}\left( q \right) \right)= \\ 
& ={{g}^{b{{b}_{1}}}}{{g}^{d{{d}_{1}}}}{{g}^{l{{l}_{1}}}}\left( \frac{\partial {{{\hat{V}}}_{bd,{{g}_{1}}{{g}_{2}}}}\left( q \right)}{\partial {{q}^{l}}}\frac{\partial {{{\hat{V}}}_{{{b}_{1}}{{d}_{1}},{{g}_{1}}{{g}_{2}}}}\left( q \right)}{\partial {{q}^{{{l}_{1}}}}} \right.- \\ 
& -ig\hat{A}_{{{l}_{1}},{{g}_{3}}}^{\left( I \right)}\left( q \right)\frac{\partial {{{\hat{V}}}_{bd,{{g}_{1}}{{g}_{2}}}}\left( q \right)}{\partial {{q}^{l}}}\hat{I}_{{{g}_{1}}{{g}_{4}}}^{{{g}_{3}}}{{{\hat{V}}}_{{{b}_{1}}{{d}_{1}},{{g}_{4}}{{g}_{2}}}}\left( q \right)-ig\hat{A}_{l,{{g}_{3}}}^{\left( II \right)}\left( q \right)\frac{\partial {{{\hat{V}}}_{bd,{{g}_{1}}{{g}_{2}}}}\left( q \right)}{\partial {{q}^{l}}}\hat{I}_{{{g}_{2}}{{g}_{4}}}^{{{g}_{3}}}{{{\hat{V}}}_{{{b}_{1}}{{d}_{1}},{{g}_{1}}{{g}_{4}}}}\left( q \right)- \\ 
& -ig\hat{A}_{l,{{g}_{3}}}^{\left( I \right)}\left( q \right)\hat{I}_{{{g}_{1}}{{g}_{4}}}^{{{g}_{3}}}{{{\hat{V}}}_{bd,{{g}_{4}}{{g}_{2}}}}\left( q \right)\frac{\partial {{{\hat{V}}}_{{{b}_{1}}{{d}_{1}},{{g}_{1}}{{g}_{2}}}}\left( q \right)}{\partial {{q}^{{{l}_{1}}}}}-ig\hat{A}_{l,{{g}_{3}}}^{\left( II \right)}\left( q \right)\hat{I}_{{{g}_{2}}{{g}_{4}}}^{{{g}_{3}}}{{{\hat{V}}}_{bd,{{g}_{1}}{{g}_{4}}}}\left( q \right)\frac{\partial {{{\hat{V}}}_{{{b}_{1}}{{d}_{1}},{{g}_{1}}{{g}_{2}}}}\left( q \right)}{\partial {{q}^{{{l}_{1}}}}} \\ 
& +{{g}^{2}}\hat{A}_{l,{{g}_{3}}}^{\left( I \right)}\left( q \right)\hat{A}_{{{l}_{1}},{{g}_{5}}}^{\left( I \right)}\left( q \right)\hat{I}_{{{g}_{4}}{{g}_{1}}}^{{{g}_{3}}}\hat{I}_{{{g}_{1}}{{g}_{6}}}^{{{g}_{5}}}{{{\hat{V}}}_{bd,{{g}_{4}}{{g}_{2}}}}\left( q \right){{{\hat{V}}}_{{{b}_{1}}{{d}_{1}},{{g}_{6}}{{g}_{2}}}}\left( q \right)+ \\ 
& +{{g}^{2}}\hat{A}_{l,{{g}_{5}}}^{\left( II \right)}\left( q \right)\hat{A}_{l,{{g}_{3}}}^{\left( II \right)}\left( q \right)\hat{I}_{{{g}_{6}}{{g}_{2}}}^{{{g}_{5}}}\hat{I}_{{{g}_{2}}{{g}_{4}}}^{{{g}_{3}}}{{{\hat{V}}}_{bd,{{g}_{1}}{{g}_{6}}}}\left( q \right){{{\hat{V}}}_{{{b}_{1}}{{d}_{1}},{{g}_{1}}{{g}_{4}}}}\left( q \right)+ \\ 
& -2{{g}^{2}}\hat{A}_{l,{{g}_{3}}}^{\left( I \right)}\left( q \right)\hat{A}_{{{l}_{1}},{{g}_{5}}}^{\left( II \right)}\left( q \right)\hat{I}_{{{g}_{1}}{{g}_{4}}}^{{{g}_{3}}}\hat{I}_{{{g}_{2}}{{g}_{6}}}^{{{g}_{5}}}{{{\hat{V}}}_{bd,{{g}_{4}}{{g}_{2}}}}\left( q \right){{{\hat{V}}}_{{{b}_{1}}{{d}_{1}},{{g}_{1}}{{g}_{6}}}}\left( q \right)\Biggr).\\
\end{split}
\end{equation}
Тут ми скористалися тим, що матричні елементи генераторів приєднаного представлення співпадають із структурними константами і, відтак є антисиметричними відносно перестановок довільної пари індексів. Тепер ми можемо не порушуючи інваріантність лагранжіану відносно локального $S{{U}_{c}}\left( 3 \right)-$ перетворення замінити добутки одночастинкових полів $ \hat{A}_{l,{{g}_{1}}}^{\left( I \right)}\left( q \right)\hat{A}_{{{l}_{1}},{{g}_{2}}}^{\left( I \right)}\left( q \right),$  $\hat{A}_{l,{{g}_{1}}}^{\left( II \right)}\left( q \right)\hat{A}_{l,{{g}_{2}}}^{\left( II \right)}\left( q \right),\hat{A}_{l,{{g}_{1}}}^{\left( I \right)}\left( q \right)\hat{A}_{{{l}_{1}},{{g}_{2}}}^{\left( II \right)}\left( q \right) $ на тензори $ \hat{A}_{l{{l}_{1}},{{g}_{1}}{{g}_{2}}}^{\left( I,I \right)}\left( q \right),\hat{A}_{l{{l}_{1}},{{g}_{1}}{{g}_{2}}}^{\left( II,II \right)}\left( q \right),\hat{A}_{l{{l}_{1}},{{g}_{1}}{{g}_{2}}}^{\left( I,II \right)}\left( q \right) $, що описують двочастинкові поля з тим самим законом перетворення, що й ці добутки. Оскільки, закон перетворення одночастинкового поля співпадає із \eqref{Zacon_odnogluonnnogo_pola}, то й закон перетворення цих тензорів співпадає з \eqref{Zacon_peretvorenna_dla_dvogluonnogo_tenzora}. Оскільки поле $\hat{V}\left( q \right),$ що входить у лагранжіан \eqref{Lagrangian_z_polem_V} є проєкцією лінійного простору тензорів $ {{\hat{V}}_{bd,{{g}_{1}}{{g}_{2}}}} $ на інваріантний  підпростір, на якому реалізуються скалярні представлення групи перетворень при переході із однієї системи відліку до іншої, і групи  $S{{U}_{c}}\left( 3 \right),$ то таку ж проєкцію зручно тепер виділити і в лагранжіані для поля $ {{\hat{V}}_{bd,{{g}_{1}}{{g}_{2}}}}\left( q\right) $. Розкладаючи лінійний простір тензорів $ {{\hat{V}}_{bd,{{g}_{1}}{{g}_{2}}}} $ у пряму суму інваріантних підпросторів запишемо:
\begin{equation}\label{Rozcladanna_v_pramu_sumu}
{{\hat{V}}_{bd,{{g}_{1}}{{g}_{2}}}}\left( q \right)=k{{g}_{bd}}{{\delta }_{{{g}_{1}}{{g}_{2}}}}\hat{V}\left( q \right)+\ldots 
\end{equation}
Тут $k-$ нормувальний коефіцієнт, а через <<трикрапки>> позначено проекції на решту інваріантних підпросторів. Оскільки ці проекції не входять у лагранжіан взаємодії \eqref{Lagrangian_z_polem_V}, то з метою отримати для початку найпростішу модель ми покладемо їх рівними нулю. Коефіцієнт $k$ знайдемо, підставляючи розклад \eqref{Rozcladanna_v_pramu_sumu} в \eqref{Poznachenna_Aplus_Aminus_V}:
\begin{equation}\label{Rivnanna_dla_k}
\hat{V}\left( q \right)=k\left( \sum\limits_{b=4}^{9}{\sum\limits_{d=4}^{9}{{{g}^{bd}}}}{{g}_{bd}}{{\delta }^{{{g}_{1}}{{g}_{2}}}}{{\delta }_{{{g}_{1}}{{g}_{2}}}} \right)\hat{V}\left( q \right)
\end{equation}
Враховуючи вид метричного тензора \eqref{Scalarnij_dobutoc_na_pidmnojini_odnochasnosti} і те, що індекси ${{g}_{1}}$ і ${{g}_{1}}$ приймають значення від 1 до 8, маємо:  
\begin{equation}\label{Kravno1_1350}
k=\frac{1}{\text{1350}}
\end{equation}
З урахуванням обговорених перетворень, лагранжіан поля $\hat{V}\left( q \right)$ прийме вид:
\begin{equation}\label{Lagrangian_pola_V_localno_invariantnij}
\begin{split}
& {{L}_{V}}=\frac{1}{2}{{g}^{l{{l}_{1}}}}k\frac{\partial \hat{V}\left( q \right)}{\partial {{q}^{l}}}\frac{\partial \hat{V}\left( q \right)}{\partial {{q}^{{{l}_{1}}}}}-\frac{k}{2}M_{G}^{2}{{{\hat{V}}}^{2}}\left( q \right)+ \\ 
& +\frac{1}{2}{{k}^{2}}{{g}^{b{{b}_{1}}}}{{g}_{{{b}_{1}}{{d}_{1}}}}{{g}_{bd}}{{g}^{d{{d}_{1}}}}{{g}^{l{{l}_{1}}}}{{g}^{2}}\hat{A}_{l{{l}_{1}},{{g}_{3}}{{g}_{5}}}^{\left( I,I \right)}\left( q \right)\hat{I}_{{{g}_{2}}{{g}_{1}}}^{{{g}_{3}}}\hat{I}_{{{g}_{1}}{{g}_{2}}}^{{{g}_{5}}}{{{\hat{V}}}^{2}}\left( q \right)+ \\ 
& +\frac{1}{2}{{k}^{2}}{{g}^{b{{b}_{1}}}}{{g}_{bd}}{{g}_{{{b}_{1}}{{d}_{1}}}}{{g}^{d{{d}_{1}}}}{{g}^{l{{l}_{1}}}}{{g}^{2}}\hat{A}_{l{{l}_{1}},{{g}_{5}}{{g}_{3}}}^{\left( II,II \right)}\left( q \right)\hat{I}_{{{g}_{1}}{{g}_{2}}}^{{{g}_{5}}}\hat{I}_{{{g}_{2}}{{g}_{1}}}^{{{g}_{3}}}{{{\hat{V}}}^{2}}\left( q \right)+ \\ 
& -{{k}^{2}}{{g}^{b{{b}_{1}}}}{{g}_{bd}}{{g}_{{{b}_{1}}{{d}_{1}}}}{{g}^{d{{d}_{1}}}}{{g}^{l{{l}_{1}}}}{{g}^{2}}\hat{A}_{l{{l}_{1}},{{g}_{3}}{{g}_{5}}}^{\left( I,II \right)}\left( q \right)\hat{I}_{{{g}_{1}}{{g}_{2}}}^{{{g}_{3}}}\hat{I}_{{{g}_{2}}{{g}_{1}}}^{{{g}_{5}}}{{{\hat{V}}}^{2}}\left( q \right). \\ 
\end{split}
\end{equation}
Одночастинкові калібрувальні поля, як і раніше було показано, не входять у лагранжіан і можуть бути покладені рівними нулю. При цьому, закон перетворення двочастинкових полів співпаде із \eqref{Zacon_peretvorenna_dla_dvogluonnogo_tenzora}.
Безпосереднім розрахунком можна переконатися, що для генераторів приєднаного представлення має місце тотожність:
\begin{equation}\label{Totognist_dla_generatoriv_priednanogo_predstavlenna}
\hat{I}_{{{g}_{1}}{{g}_{2}}}^{{{g}_{3}}}\hat{I}_{{{g}_{2}}{{g}_{1}}}^{{{g}_{4}}}=2{{\delta }^{{{g}_{3}}{{g}_{4}}}}
\end{equation}
З урахуванням цієї тотожності лагранжіан поля $ \hat{V}\left( q \right) $ приймає вид:
\begin{equation}\label{Lagrangian_pola_V_z_uraxuvannam_zgortci_generatoriv}
\begin{split}
& {{L}_{V}}=\frac{1}{2}{{g}^{l{{l}_{1}}}}k\frac{\partial \hat{V}\left( q \right)}{\partial {{q}^{l}}}\frac{\partial \hat{V}\left( q \right)}{\partial {{q}^{{{l}_{1}}}}}-\frac{k}{2}M_{G}^{2}{{{\hat{V}}}^{2}}\left( q \right)+ \\ 
& +{{g}^{2}}k{{g}^{l{{l}_{1}}}}{{\delta }^{{{g}_{3}}{{g}_{5}}}}\left( \hat{A}_{l{{l}_{1}},{{g}_{3}}{{g}_{5}}}^{\left( I,I \right)}\left( q \right)+\hat{A}_{l{{l}_{1}},{{g}_{5}}{{g}_{3}}}^{\left( II,II \right)}\left( q \right)-2\hat{A}_{l{{l}_{1}},{{g}_{3}}{{g}_{5}}}^{\left( I,II \right)}\left( q \right) \right){{{\hat{V}}}^{2}}\left( q \right) \\ 
\end{split}
\end{equation}
Поле $ \hat{A}_{l{{l}_{1}},{{g}_{3}}{{g}_{5}}}^{\left( I,I \right)}\left( q \right)+\hat{A}_{l{{l}_{1}},{{g}_{5}}{{g}_{3}}}^{\left( II,II \right)}\left( q \right)-2\hat{A}_{l{{l}_{1}},{{g}_{3}}{{g}_{5}}}^{\left( I,II \right)}\left( q \right) $, внаслідок взаємознищення неоднорідних доданків у законі \eqref{Zacon_peretvorenna_dla_dvogluonnogo_tenzora}, перетворюється за тензорним добутком двох приєднаних представлень групи  $S{{U}_{c}}\left( 3 \right)$. Але за тим самим законом перетворюється й поле $ {{\hat{V}}_{bd,{{g}_{1}}{{g}_{2}}}}\left( q \right) $. Оскільки, всі попередні міркування визначали не самі калібрувальні поля, а лише закон їх перетворення, з метою досягнення локальної $S{{U}_{c}}\left( 3 \right)$- інваріантності лагранжіану. Оскільки поле $ {{\hat{V}}_{bd,{{g}_{1}}{{g}_{2}}}}\left( q \right) $ має потрібний закон перетворення, то використовуючи його в якості поля $ \hat{A}_{l{{l}_{1}},{{g}_{3}}{{g}_{5}}}^{\left( I,I \right)}\left( q \right)+\hat{A}_{l{{l}_{1}},{{g}_{5}}{{g}_{3}}}^{\left( II,II \right)}\left( q \right)-2\hat{A}_{l{{l}_{1}},{{g}_{3}}{{g}_{5}}}^{\left( I,II \right)}\left( q \right) $ ми отримаємо для поля $\hat{V}\left( q \right)$  $S{{U}_{c}}\left( 3 \right)$ локально інваріантний лагранжіан 
\begin{equation}\label{Oconchatelnij_lagrangian_pola_V}
\begin{split}
{{L}_{V}}=\frac{1}{2}{{g}^{l{{l}_{1}}}}k\frac{\partial \hat{V}\left( q \right)}{\partial {{q}^{l}}}\frac{\partial \hat{V}\left( q \right)}{\partial {{q}^{{{l}_{1}}}}}-\frac{k}{2}M_{G}^{2}{{\hat{V}}^{2}}\left( q \right)+{{g}^{2}}k{{\hat{V}}^{3}}\left( q \right).
\end{split}
\end{equation}
Додаючи цей лагранжіан до \eqref{Lagrangian_z_polem_V} отримаємо лагранжіан динамічної моделі взаємодіючих багаточастинкових полів:
\begin{equation}\label{Povnij_Lagrangian_z_polem_V}
\begin{split}
& \frac{L}{6}=\frac{i}{2}\left( \sum\limits_{b=0}^{3}{\left( {{{\hat{\bar{\Psi }}}}_{{{s}_{1}}}}\left( q \right)\gamma _{{{s}_{1}}{{s}_{2}}}^{b}\frac{\partial {{{\hat{\Psi }}}_{{{s}_{2}}}}\left( q \right)}{\partial {{q}^{b}}}-\frac{\partial {{{\hat{\bar{\Psi }}}}_{{{s}_{1}}}}\left( q \right)}{\partial {{q}^{b}}}\gamma _{{{s}_{1}}{{s}_{2}}}^{b}{{{\hat{\Psi }}}_{{{s}_{2}}}}\left( q \right) \right)} \right)- \\ 
& -\left( 3{{m}_{q}} \right){{{\hat{\bar{\Psi }}}}_{{{s}_{1}}}}\left( q \right){{{\hat{\Psi }}}_{{{s}_{1}}}}\left( q \right)+\frac{1}{2\left( 3{{m}_{q}} \right)}\sum\limits_{b=4}^{9}{\sum\limits_{d=4}^{9}{{{g}^{bd}}\frac{\partial {{{\hat{\bar{\Psi }}}}_{{{s}_{1}}}}\left( q \right)}{\partial {{q}^{b}}}\frac{\partial {{{\hat{\Psi }}}_{{{s}_{1}}}}\left( q \right)}{\partial {{q}^{d}}}}}+ \\ 
&+\frac{{{g}^{2}}}{\left( 9{{m}_{q}} \right)}{{{\hat{\bar{\Psi }}}}_{{{s}_{1}}}}\left( q \right){{{\hat{\Psi }}}_{{{s}_{1}}}}\left( q \right)\hat{V}\left( q \right) +\\
&+\frac{1}{2}{{g}^{l{{l}_{1}}}}\frac{k}{6}\frac{\partial \hat{V}\left( q \right)}{\partial {{q}^{l}}}\frac{\partial \hat{V}\left( q \right)}{\partial {{q}^{{{l}_{1}}}}}-\frac{1}{2}\frac{k}{6}M_{G}^{2}{{\hat{V}}^{2}}\left( q \right)+{{g}^{2}}\frac{k}{6}{{\hat{V}}^{3}}\left( q \right) 
\end{split}
\end{equation}
Замість поля $ \hat{V}\left( q \right) $ введемо нове поле $ \hat{u}\left( q \right) $ згідно із співвідношенням: 
\begin{equation}\label{U_ot_q}
\hat{V}\left( q \right)=-\sqrt{\frac{6}{k}}\hat{u}\left( q \right).
\end{equation}
Обрання знаку в цьому виразі буде пояснено пізніше. Після цієї заміни отримаємо:
\begin{equation}\label{Lagrangian_z_polem_U}
\begin{split}
& \frac{L}{6}=\frac{i}{2}\left( \sum\limits_{b=0}^{3}{\left( {{{\hat{\bar{\Psi }}}}_{{{s}_{1}}}}\left( q \right)\gamma _{{{s}_{1}}{{s}_{2}}}^{b}\frac{\partial {{{\hat{\Psi }}}_{{{s}_{2}}}}\left( q \right)}{\partial {{q}^{b}}}-\frac{\partial {{{\hat{\bar{\Psi }}}}_{{{s}_{1}}}}\left( q \right)}{\partial {{q}^{b}}}\gamma _{{{s}_{1}}{{s}_{2}}}^{b}{{{\hat{\Psi }}}_{{{s}_{2}}}}\left( q \right) \right)} \right)- \\ 
& -\left( 3{{m}_{q}} \right){{{\hat{\bar{\Psi }}}}_{{{s}_{1}}}}\left( q \right){{{\hat{\Psi }}}_{{{s}_{1}}}}\left( q \right)+\frac{1}{2\left( 3{{m}_{q}} \right)}\sum\limits_{b=4}^{9}{\sum\limits_{d=4}^{9}{{{g}^{bd}}\frac{\partial {{{\hat{\bar{\Psi }}}}_{{{s}_{1}}}}\left( q \right)}{\partial {{q}^{b}}}\frac{\partial {{{\hat{\Psi }}}_{{{s}_{1}}}}\left( q \right)}{\partial {{q}^{d}}}}}+ \\ 
& -\frac{{{g}^{2}}}{\left( 9{{m}_{q}} \right)}\sqrt{\frac{6}{k}}{{{\hat{\bar{\Psi }}}}_{{{s}_{1}}}}\left( q \right){{{\hat{\Psi }}}_{{{s}_{1}}}}\left( q \right)\hat{u}\left( q \right)+ \\
&+\frac{1}{2}{{g}^{l{{l}_{1}}}}\frac{\partial \hat{u}\left( q \right)}{\partial {{q}^{l}}}\frac{\partial \hat{u}\left( q \right)}{\partial {{q}^{{{l}_{1}}}}}-\frac{1}{2}M_{G}^{2}{{\hat{u}}^{2}}\left( q \right)-{{g}^{2}}\sqrt{\frac{6}{k}}{{\hat{u}}^{3}}\left( q \right). 
\end{split}
\end{equation}
Далі зручно перейти від змінних $q$ до нових змінних $z$ згідно із співвідношенням:
\begin{equation}\label{Perehid_vid_q_do_z}
q=\left( \begin{matrix}
{{z}^{0}}  \\
{{z}^{1}}  \\
{{z}^{2}}  \\
{{z}^{3}}  \\
\sqrt{\frac{2}{9}}{{z}^{4}}  \\
\sqrt{\frac{2}{9}}{{z}^{5}}  \\
\sqrt{\frac{2}{9}}{{z}^{6}}  \\
\sqrt{\frac{1}{6}}{{z}^{7}}  \\
\sqrt{\frac{1}{6}}{{z}^{8}}  \\
\sqrt{\frac{1}{6}}{{z}^{9}}  \\
\end{matrix} \right)
\end{equation}
Після цієї заміни, лагранжіан приймає вид:
\begin{equation}\label{Lagrangian_ot_z}
\begin{split}
  & \frac{L}{6}=\frac{i}{2}\left( \sum\limits_{b=0}^{3}{\left( {{{\hat{\bar{\Psi }}}}_{{{s}_{1}}}}\left( z \right)\gamma _{{{s}_{1}}{{s}_{2}}}^{b}\frac{\partial {{{\hat{\Psi }}}_{{{s}_{2}}}}\left( z \right)}{\partial {{z}^{b}}}-\frac{\partial {{{\hat{\bar{\Psi }}}}_{{{s}_{1}}}}\left( z \right)}{\partial {{z}^{b}}}\gamma _{{{s}_{1}}{{s}_{2}}}^{b}{{{\hat{\Psi }}}_{{{s}_{2}}}}\left( z \right) \right)} \right)- \\ 
& -\left( \frac{1}{2\left( 3{{m}_{q}} \right)}\sum\limits_{b=4}^{10}{\frac{\partial {{{\hat{\bar{\Psi }}}}_{{{s}_{1}}}}\left( z \right)}{\partial {{z}^{b}}}\frac{\partial {{{\hat{\Psi }}}_{{{s}_{1}}}}\left( z \right)}{\partial {{z}^{b}}}}+ \right. \\ 
& \left. +\frac{{{g}^{2}}}{\left( 9{{m}_{q}} \right)}\sqrt{\frac{6}{k}}{{{\hat{\bar{\Psi }}}}_{{{s}_{1}}}}\left( z \right){{{\hat{\Psi }}}_{{{s}_{1}}}}\left( z \right)\hat{u}\left( z \right)+\left( 3{{m}_{q}} \right){{{\hat{\bar{\Psi }}}}_{{{s}_{1}}}}\left( z \right){{{\hat{\Psi }}}_{{{s}_{1}}}}\left( z \right) \right)+ \\ 
& +\frac{1}{2}\sum\limits_{{{l}_{1}}=0}^{3}{\sum\limits_{{{l}_{2}}=0}^{3}{{{g}^{{{l}_{1}}{{l}_{2}}}}\frac{\partial \hat{u}\left( z \right)}{\partial {{z}^{{{l}_{1}}}}}\frac{\partial \hat{u}\left( z \right)}{\partial {{z}^{{{l}_{2}}}}}}}-\frac{1}{2}\sum\limits_{l=4}^{9}{{{\left( \frac{\partial \hat{u}\left( z \right)}{\partial {{z}^{l}}} \right)}^{2}}}-\frac{1}{2}M_{G}^{2}{{{\hat{u}}}^{2}}\left( z \right)-{{g}^{2}}\sqrt{\frac{6}{k}}{{{\hat{u}}}^{3}}\left( z \right). \\ 
\end{split}
\end{equation}
Для подальшого аналізу зручно перейти до безрозмірних величин. У якості характерної маси природно обрати масу протона, яку позначатимемо ${{M}_{P}}.$ Тоді, характерна довжина $M_{P}^{-1}.$ Оскільки дія безрозмірна, лагранжіан має розмірність $M_{P}^{10}.$ Введемо далі безрозмірні параметри ${{\mu }_{q}}$ і ${{m}_{G}}:$
\begin{equation}\label{Bezrozmirni_masi}
{{\mu }_{q}}=\frac{{{m}_{q}}}{{{M}_{P}}},{{m}_{G}}=\frac{{{M}_{G}}}{{{M}_{P}}}.
\end{equation} 
Введемо також безрозмірні координати ${{\rho }^{a}}$ і безрозмірні поля ${{\hat{\bar{\psi }}}_{{{s}_{1}}}}\left( \rho  \right),{{\hat{\psi }}_{{{s}_{1}}}}\left( \rho  \right),\hat{v}\left( \rho  \right)$:
\begin{equation}\label{Bezrozmirni_coordinati_i_pola}
{{\rho }^{a}}={{M}_{P}}{{z}^{a}},{{\hat{\bar{\Psi }}}_{{{s}_{1}}}}\left( \rho  \right)=M_{P}^{{{n}_{\psi }}}{{\hat{\bar{\psi }}}_{{{s}_{1}}}}\left( \rho  \right),{{\hat{\Psi }}_{{{s}_{1}}}}\left( \rho  \right)=M_{P}^{{{n}_{\psi }}}{{\hat{\psi }}_{{{s}_{1}}}}\left( \rho  \right),\hat{u}\left( \rho  \right)=M_{P}^{{{n}_{u}}}\hat{v}\left( \rho  \right).
\end{equation}
Тут ${n}_{\psi }$ і $ {n}_{u} $ - поки що невідомі ступені, які ми збираємось знайти з умови, щоб усі доданки лагранжіану мали однакову розмірність. Після замін \eqref{Bezrozmirni_masi} і \eqref{Bezrozmirni_coordinati_i_pola} лагранжіан прийме вид:
\begin{equation}\label{Lagrangian_obezrazmerivanie  }
\begin{split}
  & \frac{L}{6}=\frac{i}{2}\left( M_{P}^{2{{n}_{\psi }}+1}\sum\limits_{b=0}^{3}{\left( {{{\hat{\bar{\psi }}}}_{{{s}_{1}}}}\left( \rho  \right)\gamma _{{{s}_{1}}{{s}_{2}}}^{b}\frac{\partial {{{\hat{\psi }}}_{{{s}_{2}}}}\left( \rho  \right)}{\partial {{\rho }^{b}}}-\frac{\partial {{{\hat{\bar{\psi }}}}_{{{s}_{1}}}}\left( \rho  \right)}{\partial {{\rho }^{b}}}\gamma _{{{s}_{1}}{{s}_{2}}}^{b}{{{\hat{\psi }}}_{{{s}_{2}}}}\left( \rho  \right) \right)} \right)- \\ 
& -\left( \frac{1}{2\left( 3{{\mu }_{q}} \right)}M_{P}^{2{{n}_{\psi }}+1}\sum\limits_{b=4}^{10}{\frac{\partial {{{\hat{\bar{\psi }}}}_{{{s}_{1}}}}\left( \rho  \right)}{\partial {{\rho }^{b}}}\frac{\partial {{{\hat{\psi }}}_{{{s}_{1}}}}\left( \rho  \right)}{\partial {{\rho }^{b}}}}+ \right. \\ 
& \left. +M_{P}^{2{{n}_{\psi }}+{{n}_{u}}-1}\frac{{{g}^{2}}}{\left( 9{{\mu }_{q}} \right)}\sqrt{\frac{6}{k}}{{{\hat{\bar{\psi }}}}_{{{s}_{1}}}}\left( \rho  \right){{{\hat{\psi }}}_{{{s}_{1}}}}\left( \rho  \right)\hat{v}\left( \rho  \right)+\left( 3{{\mu }_{q}} \right)M_{P}^{2{{n}_{\psi }}+1}{{{\hat{\bar{\psi }}}}_{{{s}_{1}}}}\left( \rho  \right){{{\hat{\psi }}}_{{{s}_{1}}}}\left( \rho  \right) \right)+ \\ 
& +\frac{1}{2}M_{P}^{2{{n}_{u}}+2}\sum\limits_{{{l}_{1}}=0}^{3}{\sum\limits_{{{l}_{2}}=0}^{3}{{{g}^{{{l}_{1}}{{l}_{2}}}}\frac{\partial \hat{v}\left( \rho  \right)}{\partial {{\rho }^{{{l}_{1}}}}}\frac{\partial \hat{v}\left( \rho  \right)}{\partial {{\rho }^{{{l}_{2}}}}}}}-\frac{1}{2}M_{P}^{2{{n}_{u}}+2}\sum\limits_{l=4}^{9}{{{\left( \frac{\partial \hat{v}\left( \rho  \right)}{\partial {{\rho }^{l}}} \right)}^{2}}}- \\ 
& -\frac{1}{2}m_{G}^{2}M_{P}^{2{{n}_{u}}+2}{{{\hat{v}}}^{2}}\left( \rho  \right)-{{g}^{2}}M_{P}^{3{{n}_{u}}}\sqrt{\frac{6}{k}}{{{\hat{v}}}^{3}}\left( \rho  \right). \\ 
\end{split}
\end{equation}
З цього виразу видно, що для того, щоб усі доданки лагранжіану мали спільний множник, потрібно
\begin{equation}\label{n_u_n_psi}
{{n}_{u}}=2,{{n}_{\psi }}=2.5.
\end{equation}
Тоді, лагранжіан матиме розмірність $M_{P}^{6}.$ Для досягнення потрібної розмірності весь лагранжіан ми повинні множити на $M_{P}^{4}.$ Оскільки, нас далі цікавитимуть лише динамічні рівняння, то зручно буде розглядати безрозмірний лагранжіан:
\begin{equation}\label{Bezrozmirnij_lagrangian}
\begin{split}
  & l=\frac{i}{2}\left( \sum\limits_{b=0}^{3}{\left( {{{\hat{\bar{\psi }}}}_{{{s}_{1}}}}\left( \rho  \right)\gamma _{{{s}_{1}}{{s}_{2}}}^{b}\frac{\partial {{{\hat{\psi }}}_{{{s}_{2}}}}\left( \rho  \right)}{\partial {{\rho }^{b}}}-\frac{\partial {{{\hat{\bar{\psi }}}}_{{{s}_{1}}}}\left( \rho  \right)}{\partial {{\rho }^{b}}}\gamma _{{{s}_{1}}{{s}_{2}}}^{b}{{{\hat{\psi }}}_{{{s}_{2}}}}\left( \rho  \right) \right)} \right)- \\ 
& -\left( \frac{1}{2\left( 3{{\mu }_{q}} \right)}\sum\limits_{b=4}^{10}{\frac{\partial {{{\hat{\bar{\psi }}}}_{{{s}_{1}}}}\left( \rho  \right)}{\partial {{\rho }^{b}}}\frac{\partial {{{\hat{\psi }}}_{{{s}_{1}}}}\left( \rho  \right)}{\partial {{\rho }^{b}}}}+ \right. \\ 
& \left. +\frac{{{g}^{2}}}{\left( 9{{\mu }_{q}} \right)}\sqrt{\frac{6}{k}}{{{\hat{\bar{\psi }}}}_{{{s}_{1}}}}\left( \rho  \right){{{\hat{\psi }}}_{{{s}_{1}}}}\left( \rho  \right)\hat{v}\left( \rho  \right)+\left( 3{{\mu }_{q}} \right){{{\hat{\bar{\psi }}}}_{{{s}_{1}}}}\left( \rho  \right){{{\hat{\psi }}}_{{{s}_{1}}}}\left( \rho  \right) \right)+ \\ 
& +\frac{1}{2}\sum\limits_{{{l}_{1}}=0}^{3}{\sum\limits_{{{l}_{2}}=0}^{3}{{{g}^{{{l}_{1}}{{l}_{2}}}}\frac{\partial \hat{v}\left( \rho  \right)}{\partial {{\rho }^{{{l}_{1}}}}}\frac{\partial \hat{v}\left( \rho  \right)}{\partial {{\rho }^{{{l}_{2}}}}}}}-\frac{1}{2}\sum\limits_{l=4}^{9}{{{\left( \frac{\partial \hat{v}\left( \rho  \right)}{\partial {{\rho }^{l}}} \right)}^{2}}}-\frac{1}{2}m_{G}^{2}{{{\hat{v}}}^{2}}\left( \rho  \right)-{{g}^{2}}\sqrt{\frac{6}{k}}{{{\hat{v}}}^{3}}\left( \rho  \right). \\ 
\end{split}
\end{equation}
Для подальшого розгляду зручно виділити безрозмірні координати центру мас і безрозмірні внутрішні координати. Тому, введемо позначення:
\begin{equation}\label{Bezrpzmirni_ro_X_y}
	\begin{split}
 & \rho _{X}^{b}={{\rho }^{b}},b=0,1,2,3, \\ 
& \rho _{y}^{b}={{\rho }^{b}},b=4,5,\ldots ,9. \\ 		
	\end{split}
\end{equation}
Для цих змінних також використовуватимемо позначення ${{\rho }_{X}}$ i ${{\rho }_{y}}$, коли мова йтиме про всю сукупність тих або інших безрозмірних координат. Нехай ${{v}_{0}}\left( {{\rho }_{y}} \right)-$ деяка функція із числовими значеннями від внутрішніх координат. Представимо поле $ \hat{v}\left( \rho  \right) $ у виді: 
\begin{equation}\label{v0_plus_v1}
\hat{v}\left( \rho  \right)={{v}_{0}}\left( {{\rho }_{y}} \right)\hat{E}+{{\hat{v}}_{1}}\left( \rho  \right)
\end{equation}
Тут $ {{\hat{v}}_{1}}\left( \rho  \right)-$ нова динамічна змінна (операторнозначна польова функція), а $ \hat{E}- $ одиничний оператор. Представимо також тричастинкове біспінорне поле в виді:
\begin{equation}\label{Razdelrnie_bisapinornogo_pola}
{{\hat{\psi }}_{{{s}_{1}}}}\left( \rho  \right)={{\hat{\Psi }}_{{{s}_{1}}}}\left( {{\rho }_{X}} \right)\phi \left( {{\rho }_{y}} \right),{{\hat{\bar{\psi }}}_{{{s}_{1}}}}\left( \rho  \right)={{\hat{\bar{\Psi }}}_{{{s}_{1}}}}\left( {{\rho }_{X}} \right){{\phi }^{*}}\left( {{\rho }_{y}} \right)
\end{equation}
Тут $ {{\hat{\bar{\Psi }}}_{{{s}_{1}}}}\left( {{\rho }_{X}} \right),{{\hat{\Psi }}_{{{s}_{1}}}}\left( {{\rho }_{X}} \right)-$ також нові операторнозначні польові функції, а $ {{\phi }^{*}}\left( {{\rho }_{y}} \right),\phi \left( {{\rho }_{y}} \right)-$взаємно комплексно спряжені функції, що приймають числові значення. Використовуючи \eqref{v0_plus_v1} і \eqref{Razdelrnie_bisapinornogo_pola}, а також інтегруючи по частинах по внутрішніх координатах, лагранжіан \eqref{Bezrozmirnij_lagrangian} може бути переписаний у виді:
\begin{equation}\label{Bezrozmirnij_lagrangian_posle_v0_plus_v1}
	\begin{split}
    & l=\frac{i}{2}\left( \sum\limits_{b=0}^{3}{\left( {{{\hat{\bar{\Psi }}}}_{{{s}_{1}}}}\left( {{\rho }_{X}} \right)\gamma _{{{s}_{1}}{{s}_{2}}}^{b}\frac{\partial {{{\hat{\Psi }}}_{{{s}_{2}}}}\left( {{\rho }_{X}} \right)}{\partial \rho _{X}^{b}}-\frac{\partial {{{\hat{\bar{\psi }}}}_{{{s}_{1}}}}\left( {{\rho }_{X}} \right)}{\partial \rho _{X}^{b}}\gamma _{{{s}_{1}}{{s}_{2}}}^{b}{{{\hat{\psi }}}_{{{s}_{2}}}}\left( {{\rho }_{X}} \right) \right)} \right){{\phi }^{*}}\left( {{\rho }_{y}} \right)\phi \left( {{\rho }_{y}} \right)- \\ 
  & -{{\phi }^{*}}\left( {{\rho }_{y}} \right)\left( -\frac{1}{2\left( 3{{\mu }_{q}} \right)}\sum\limits_{b=4}^{10}{\frac{{{\partial }^{2}}\phi \left( {{\rho }_{y}} \right)}{{{\left( \partial \rho _{y}^{b} \right)}^{2}}}+\frac{{{g}^{2}}}{\left( 9{{\mu }_{q}} \right)}\sqrt{\frac{6}{k}}{{v}_{0}}\left( {{\rho }_{y}} \right)\phi \left( {{\rho }_{y}} \right)+\left( 3{{\mu }_{q}} \right)\varphi \left( {{\rho }_{y}} \right)} \right)\times  \\ 
  & \times {{{\hat{\bar{\Psi }}}}_{{{s}_{1}}}}\left( {{\rho }_{X}} \right){{{\hat{\Psi }}}_{{{s}_{1}}}}\left( {{\rho }_{X}} \right)+ \\ 
  & +\frac{1}{2}\sum\limits_{{{l}_{1}}=0}^{3}{\sum\limits_{{{l}_{2}}=0}^{3}{{{g}^{{{l}_{1}}{{l}_{2}}}}\frac{\partial {{{\hat{v}}}_{1}}\left( \rho  \right)}{\partial {{\rho }^{{{l}_{1}}}}}\frac{\partial {{{\hat{v}}}_{1}}\left( \rho  \right)}{\partial {{\rho }^{{{l}_{2}}}}}}}-\frac{1}{2}\sum\limits_{l=4}^{9}{{{\left( \frac{\partial {{v}_{0}}\left( {{\rho }_{y}} \right)}{\partial {{\rho }^{l}}}\hat{E}+\frac{\partial {{{\hat{v}}}_{1}}\left( \rho  \right)}{\partial {{\rho }^{l}}} \right)}^{2}}} \\ 
  & -\frac{1}{2}m_{G}^{2}{{\left( {{v}_{0}}\left( {{\rho }_{y}} \right)\hat{E}+{{{\hat{v}}}_{1}}\left( \rho  \right) \right)}^{2}}-{{g}^{2}}\sqrt{\frac{6}{k}}{{\left( {{v}_{0}}\left( {{\rho }_{y}} \right)\hat{E}+{{{\hat{v}}}_{1}}\left( \rho  \right) \right)}^{3}}- \\ 
  & -\frac{{{g}^{2}}}{\left( 9{{\mu }_{q}} \right)}\sqrt{\frac{6}{k}}{{{\hat{\bar{\Psi }}}}_{{{s}_{1}}}}\left( {{\rho }_{X}} \right){{{\hat{\Psi }}}_{{{s}_{1}}}}\left( {{\rho }_{X}} \right){{\phi }^{*}}\left( {{\rho }_{y}} \right)\phi \left( {{\rho }_{y}} \right){{{\hat{v}}}_{1}}\left( \rho  \right). \\  
	\end{split}
\end{equation} 
Цей лагранжіан можна представити як суму трьох доданків: лагранжіану тричастинковго біспінорного поля:
\begin{equation}\label{L_psi}
	\begin{split}
		 & {{l}_{\Psi }}=\frac{i}{2}\left( \sum\limits_{b=0}^{3}{\left( {{{\hat{\bar{\Psi }}}}_{{{s}_{1}}}}\left( {{\rho }_{X}} \right)\gamma _{{{s}_{1}}{{s}_{2}}}^{b}\frac{\partial {{{\hat{\Psi }}}_{{{s}_{2}}}}\left( {{\rho }_{X}} \right)}{\partial \rho _{X}^{b}}-\frac{\partial {{{\hat{\bar{\psi }}}}_{{{s}_{1}}}}\left( {{\rho }_{X}} \right)}{\partial \rho _{X}^{b}}\gamma _{{{s}_{1}}{{s}_{2}}}^{b}{{{\hat{\psi }}}_{{{s}_{2}}}}\left( {{\rho }_{X}} \right) \right)} \right){{\phi }^{*}}\left( {{\rho }_{y}} \right)\phi \left( {{\rho }_{y}} \right)- \\ 
		& -{{\phi }^{*}}\left( {{\rho }_{y}} \right)\left( -\frac{1}{2\left( 3{{\mu }_{q}} \right)}\sum\limits_{b=4}^{10}{\frac{{{\partial }^{2}}\phi \left( {{\rho }_{y}} \right)}{{{\left( \partial \rho _{y}^{b} \right)}^{2}}}+\frac{{{g}^{2}}}{\left( 9{{\mu }_{q}} \right)}\sqrt{\frac{6}{k}}{{v}_{0}}\left( {{\rho }_{y}} \right)\phi \left( {{\rho }_{y}} \right)+\left( 3{{\mu }_{q}} \right)\varphi \left( {{\rho }_{y}} \right)} \right)\times  \\ 
		& \times {{{\hat{\bar{\Psi }}}}_{{{s}_{1}}}}\left( {{\rho }_{X}} \right){{{\hat{\Psi }}}_{{{s}_{1}}}}\left( {{\rho }_{X}} \right), \\ 
		\end{split}
\end{equation} 
лагранжіану двочастинкового калібрувального поля:
\begin{equation}\label{L_v}
	\begin{split}
		  &{{l}_{v}}=\frac{1}{2}\sum\limits_{{{l}_{1}}=0}^{3}{\sum\limits_{{{l}_{2}}=0}^{3}{{{g}^{{{l}_{1}}{{l}_{2}}}}\frac{\partial {{{\hat{v}}}_{1}}\left( \rho  \right)}{\partial {{\rho }^{{{l}_{1}}}}}\frac{\partial {{{\hat{v}}}_{1}}\left( \rho  \right)}{\partial {{\rho }^{{{l}_{2}}}}}}}-\frac{1}{2}\sum\limits_{l=4}^{9}{{{\left( \frac{\partial {{v}_{0}}\left( {{\rho }_{y}} \right)}{\partial {{\rho }^{l}}}\hat{E}+\frac{\partial {{{\hat{v}}}_{1}}\left( \rho  \right)}{\partial {{\rho }^{l}}} \right)}^{2}}} \\ 
		& -\frac{1}{2}m_{G}^{2}{{\left( {{v}_{0}}\left( {{\rho }_{y}} \right)\hat{E}+{{{\hat{v}}}_{1}}\left( \rho  \right) \right)}^{2}}-{{g}^{2}}\sqrt{\frac{6}{k}}{{\left( {{v}_{0}}\left( {{\rho }_{y}} \right)\hat{E}+{{{\hat{v}}}_{1}}\left( \rho  \right) \right)}^{3}}, \\ 
	\end{split}
\end{equation}
і лагранжіану їх взаємодії:
\begin{equation}\label{L_int}
	{{l}_{\operatorname{int}}}=-\frac{{{g}^{2}}}{\left( 9{{\mu }_{q}} \right)}\sqrt{\frac{6}{k}}{{\hat{\bar{\Psi }}}_{{{s}_{1}}}}\left( {{\rho }_{X}} \right){{\hat{\Psi }}_{{{s}_{1}}}}\left( {{\rho }_{X}} \right){{\phi }^{*}}\left( {{\rho }_{y}} \right)\phi \left( {{\rho }_{y}} \right){{\hat{v}}_{1}}\left( \rho  \right).
\end{equation}
Далі, ми можемо розглядати динаміку системи полів із лагранжіаном \eqref{Bezrozmirnij_lagrangian_posle_v0_plus_v1} в представленні взаємодії відносно лагранжіану ${{l}_{\operatorname{int}}}.$ Тобто, польові оператори можна розглядати як розв'язки динамічних рівнянь для лагранжіану, що є сумою $ {{l}_{\Psi }}+{{l}_{v}}$, а динаміку стану системи розглядати як таку, що визначається гамільтоніаном, породжуваним лагранжіаном взаємодії ${{l}_{\operatorname{int}}}.$  Це дає нам змогу розглядати окремо динаміку операторів, що породжується лагранжіаном $ {{l}_{\Psi }}$ окремо від динаміки операторів, що породжується лагранжіаном $ {{l}_{v}}.$ 

Розглянемо спочатку лагранжіан $ {{l}_{\Psi }}.$ Його можна переписати в виді:
\begin{equation}\label{l_psi_z_Hinternal}
\begin{split}
& {{l}_{\Psi }}=\frac{i}{2}\left( \sum\limits_{b=0}^{3}{\left( {{{\hat{\bar{\Psi }}}}_{{{s}_{1}}}}\left( {{\rho }_{X}} \right)\gamma _{{{s}_{1}}{{s}_{2}}}^{b}\frac{\partial {{{\hat{\Psi }}}_{{{s}_{2}}}}\left( {{\rho }_{X}} \right)}{\partial \rho _{X}^{b}}-\frac{\partial {{{\hat{\bar{\Psi }}}}_{{{s}_{1}}}}\left( {{\rho }_{X}} \right)}{\partial \rho _{X}^{b}}\gamma _{{{s}_{1}}{{s}_{2}}}^{b}{{{\hat{\Psi }}}_{{{s}_{2}}}}\left( {{\rho }_{X}} \right) \right)} \right){{\phi }^{*}}\left( {{\rho }_{y}} \right)\phi \left( {{\rho }_{y}} \right)- \\ 
& -{{\phi }^{*}}\left( {{\rho }_{y}} \right){{{\hat{H}}}^{\operatorname{i}\text{nternal}}}\left( \phi \left( {{\rho }_{y}} \right) \right){{{\hat{\bar{\Psi }}}}_{{{s}_{1}}}}\left( {{\rho }_{X}} \right){{{\hat{\Psi }}}_{{{s}_{2}}}}\left( {{\rho }_{X}} \right), \\ 
\end{split}
\end{equation}
де введено внутрішній гамільтоніан $ {{\hat{H}}^{\text{internal}}}$ тричастинкової системи:
\begin{equation}\label{sam_vnutrihnij_hamiltonian}
{{\hat{H}}^{\text{іnternal}}}\left( \phi \left( {{\rho }_{y}} \right) \right)\equiv -\frac{1}{2\left( 3{{\mu }_{q}} \right)}\sum\limits_{b=4}^{10}{\frac{{{\partial }^{2}}\phi \left( {{\rho }_{y}} \right)}{{{\left( \partial \rho _{y}^{b} \right)}^{2}}}+\frac{{{g}^{2}}}{\left( 9{{\mu }_{q}} \right)}\sqrt{\frac{6}{k}}{{v}_{0}}\left( {{\rho }_{y}} \right)\phi \left( {{\rho }_{y}} \right)+\left( 3{{\mu }_{q}} \right)\varphi \left( {{\rho }_{y}} \right)}.
\end{equation}
Як видно з цього виразу, функція $ \left( {{{g}^{2}}}/{\left( 9{{\mu }_{q}} \right)}\;\sqrt{{6}/{k}\;} \right){{v}_{0}}\left( {{\rho }_{y}} \right) $ грає роль потенційної енергії взаємодії між кварками. Відповідно функцію $ \varphi \left( {{\rho }_{y}} \right) $ можна розглядати як координатну частину внутрішнього стану баріону. Оскільки, залежність польових функцій від їх аргументів у обраному представленні взаємодії повинна виражати динаміку вільного баріону, то функція $ \varphi \left( {{\rho }_{y}} \right) $ повинна бути власною функцією внутрішнього гамільтоніану \eqref{sam_vnutrihnij_hamiltonian}, яка відповідає найменшому власному значенню, яке дорівнює масі баріону. Маючи на меті опис протону, покладемо її рівній масі протону ${{M}_{P}}$:
\begin{equation}\label{zadacha_na_vlasni_znachenna}
-\frac{1}{2\left( 3{{\mu }_{q}} \right)}\sum\limits_{b=4}^{10}{\frac{{{\partial }^{2}}\phi \left( {{\rho }_{y}} \right)}{{{\left( \partial \rho _{y}^{b} \right)}^{2}}}+\frac{{{g}^{2}}}{\left( 9{{\mu }_{q}} \right)}\sqrt{\frac{6}{k}}{{v}_{0}}\left( {{\rho }_{y}} \right)\phi \left( {{\rho }_{y}} \right)+\left( 3{{\mu }_{q}} \right)\varphi \left( {{\rho }_{y}} \right)}={{M}_{P}}\varphi \left( {{\rho }_{y}} \right).
\end{equation}
Якщо власну функцію $ \phi \left( {{\rho }_{y}} \right)$ обрати нормованою на одиницю, то дія для лагранжіану ${{l}_{\Psi }}$ зведеться до звичайного виразу для біспінорного поля, а динамічні рівняння - до системи рівнянь Дірака. Відповідно, матимемо загальний розв'язок цієї системи в виді лінійної комбінації негативно-частотних і позитивно-частотних розв'язків. При цьому, негативно- і позитивно-частотні коефіцієнти при реалізації звичайної процедури квантування \cite{Bogolyubov_rus} будуть описувати народження і знищення частинок із масою протону і спіном 1/2 з внутрішнім станом, що перетворюється за тривіальним представленням групи $S{{U}_{c}}\left( 3 \right)$. Окрім того, оскільки нас цікавить лише сильна взаємодія, ми не розглядали явно ароматовий стан системи трьох кварків, але він може бути обраний таким, що відповідає стану протону. Тобто, оператори народження і знищення, які відповідають полю $ {{\hat{\Psi }}_{s}}\left( {{\rho }_{X}} \right),s=1,2,3,4 $ описують процеси народження і знищення протонів. 

Розглянемо тепер лагранжіан $ {{l}_{v}}$ \eqref{L_v}. З виду цього лагранжіану бачимо, що динамічні рівняння Лагранжа - Ейлера можна отримати для поля $ \hat{v}\left( \rho  \right)={{v}_{0}}\left( {{\rho }_{y}} \right)\hat{E}+{{\hat{v}}_{1}}\left( \rho  \right) $ \eqref{v0_plus_v1}. Це означає, що функція $ {{v}_{0}}\left( {{\rho }_{y}} \right) $ буде частковим розв'язком цих рівнянь, якщо $ {{\hat{v}}_{1}}\left( \rho  \right)=0. $ Тоді представлення $ \hat{v}\left( \rho  \right)={{v}_{0}}\left( {{\rho }_{y}} \right)\hat{E}+{{\hat{v}}_{1}}\left( \rho  \right) $ можна розглядати таким чином, що ми маємо <<великий розв'язок>>  ${{v}_{0}}\left( {{\rho }_{y}} \right)$, який описує взаємодію кварків всередині баріону і квантуємо малі флуктуації навколо нього $ {{\hat{v}}_{1}}\left( \rho  \right).$ Виходячи з цього, для функції ${{v}_{0}}\left( {{\rho }_{y}} \right)$ за допомогою лагранжіану \eqref{L_v} отримуємо динамічне рівняння:
\begin{equation}\label{rivnanna_dla_v0}
\sum\limits_{l=4}^{9}{\frac{{{\partial }^{2}}{{v}_{0}}\left( {{\rho }_{y}} \right)}{{{\left( \partial \rho _{y}^{l} \right)}^{2}}}}-m_{G}^{2}{{v}_{0}}\left( {{\rho }_{y}} \right)-3{{g}^{2}}\sqrt{\frac{6}{k}}{{\left( {{v}_{0}}\left( {{\rho }_{y}} \right) \right)}^{2}}=0.
\end{equation} 
Динаміка флуктуацій ${{\hat{v}}_{1}}\left( \rho  \right)$ з урахуванням \eqref{rivnanna_dla_v0} описується лагранжіаном:
\begin{equation}\label{l_v_1}
\begin{split}
& {{l}_{{{v}_{1}}}}=\frac{1}{2}\sum\limits_{{{l}_{1}}=0}^{3}{\sum\limits_{{{l}_{2}}=0}^{3}{{{g}^{{{l}_{1}}{{l}_{2}}}}\frac{\partial {{{\hat{v}}}_{1}}\left( \rho  \right)}{\partial {{\rho }^{{{l}_{1}}}}}\frac{\partial {{{\hat{v}}}_{1}}\left( \rho  \right)}{\partial {{\rho }^{{{l}_{2}}}}}}}-\frac{1}{2}\sum\limits_{l=4}^{9}{{{\left( \frac{\partial {{{\hat{v}}}_{1}}\left( \rho  \right)}{\partial {{\rho }^{l}}} \right)}^{2}}}- \\ 
& -\frac{1}{2}m_{G}^{2}{{\left( {{{\hat{v}}}_{1}}\left( \rho  \right) \right)}^{2}}-3{{g}^{2}}\sqrt{\frac{6}{k}}\left( {{v}_{0}}\left( {{\rho }_{y}} \right) \right){{\left( {{{\hat{v}}}_{1}}\left( \rho  \right) \right)}^{2}}-{{g}^{2}}\sqrt{\frac{6}{k}}{{\left( {{{\hat{v}}}_{1}}\left( \rho  \right) \right)}^{3}}. \\ 
\end{split}
\end{equation}
Кубічний доданок $ -{{g}^{2}}\sqrt{{6}/{k}\;}{{\left( {{{\hat{v}}}_{1}}\left( \rho  \right) \right)}^{3}} $ ми можемо додати до лагранжіану взаємодії \eqref{L_int}. Тоді, враховуючи, що ми використовуємо представлення взаємодії відносно лагранжіану взаємодії:
\begin{equation}\label{L_int_1}
l_{\operatorname{int}}^{1}=-\frac{{{g}^{2}}}{\left( 9{{\mu }_{q}} \right)}\sqrt{\frac{6}{k}}{{\hat{\bar{\Psi }}}_{{{s}_{1}}}}\left( {{\rho }_{X}} \right){{\hat{\Psi }}_{{{s}_{1}}}}\left( {{\rho }_{X}} \right){{\phi }^{*}}\left( {{\rho }_{y}} \right)\phi \left( {{\rho }_{y}} \right){{\hat{v}}_{1}}\left( \rho  \right)-{{g}^{2}}\sqrt{\frac{6}{k}}{{\left( {{{\hat{v}}}_{1}}\left( \rho  \right) \right)}^{3}},
\end{equation}
залежність від своїх аргументів поля $ {{\hat{v}}_{1}}\left( \rho  \right) $ буде описуватись лагранжіаном:
 \begin{equation}\label{l^0_v_1}
 \begin{split}
 & l_{{{v}_{1}}}^{\left( 0 \right)}=\frac{1}{2}\sum\limits_{{{l}_{1}}=0}^{3}{\sum\limits_{{{l}_{2}}=0}^{3}{{{g}^{{{l}_{1}}{{l}_{2}}}}\frac{\partial {{{\hat{v}}}_{1}}\left( \rho  \right)}{\partial {{\rho }^{{{l}_{1}}}}}\frac{\partial {{{\hat{v}}}_{1}}\left( \rho  \right)}{\partial {{\rho }^{{{l}_{2}}}}}}}-\frac{1}{2}\sum\limits_{l=4}^{9}{{{\left( \frac{\partial {{{\hat{v}}}_{1}}\left( \rho  \right)}{\partial {{\rho }^{l}}} \right)}^{2}}}- \\ 
 & -\frac{1}{2}m_{G}^{2}{{\left( {{{\hat{v}}}_{1}}\left( \rho  \right) \right)}^{2}}-3{{g}^{2}}\sqrt{\frac{6}{k}}\left( {{v}_{0}}\left( {{\rho }_{y}} \right) \right){{\left( {{{\hat{v}}}_{1}}\left( \rho  \right) \right)}^{2}}. \\ 
 \end{split}
 \end{equation}
За допомогою інтегрування по частинах, його можна переписати в виді:
\begin{equation}\label{l_v_s_vnutrichnim_hamiltonianom}
\begin{split}
& l_{{{v}_{1}}}^{\left( 0 \right)}=\frac{1}{2}\sum\limits_{{{l}_{1}}=0}^{3}{\sum\limits_{{{l}_{2}}=0}^{3}{{{g}^{{{l}_{1}}{{l}_{2}}}}\frac{\partial {{{\hat{v}}}_{1}}\left( \rho  \right)}{\partial {{\rho }^{{{l}_{1}}}}}\frac{\partial {{{\hat{v}}}_{1}}\left( \rho  \right)}{\partial {{\rho }^{{{l}_{2}}}}}}}- \\ 
& -{{{\hat{v}}}_{1}}\left( \rho  \right)\left( -\frac{1}{2}\sum\limits_{l=4}^{9}{\left( \frac{{{\partial }^{2}}{{{\hat{v}}}_{1}}\left( \rho  \right)}{\partial {{\left( {{\rho }^{l}} \right)}^{2}}} \right)+\frac{1}{2}m_{G}^{2}{{{\hat{v}}}_{1}}\left( \rho  \right)+3{{g}^{2}}\sqrt{\frac{6}{k}}\left( {{v}_{0}}\left( {{\rho }_{y}} \right) \right){{{\hat{v}}}_{1}}\left( \rho  \right)} \right).
\end{split}
\end{equation}
Вираз 
\begin{equation}\label{Hv_internal}
\hat{H}_{v}^{\text{internal}}\left( {{{\hat{v}}}_{1}}\left( \rho  \right) \right)=-\frac{1}{2}\sum\limits_{l=4}^{9}{\left( \frac{{{\partial }^{2}}{{{\hat{v}}}_{1}}\left( \rho  \right)}{\partial {{\left( {{\rho }^{l}} \right)}^{2}}} \right)+\frac{1}{2}m_{G}^{2}{{{\hat{v}}}_{1}}\left( \rho  \right)+3{{g}^{2}}\sqrt{\frac{6}{k}}\left( {{v}_{0}}\left( {{\rho }_{y}} \right) \right){{{\hat{v}}}_{1}}\left( \rho  \right)}
\end{equation}
формально має вид результату дії внутрішнього гамільтоніану тричастинкової системи з потенційною енергією $ 3{{g}^{2}}\left( {6}/{k}\; \right){{v}_{0}}\left( {{\rho }_{y}} \right) .$ Тому, представляючи поле $ {{\hat{v}}_{1}}\left( \rho  \right) $ в виді: 
\begin{equation}\label{Razlojenie_v1_ro}
{{\hat{v}}_{1}}\left( \rho  \right)={{\hat{V}}_{1}}\left( {{\rho }_{X}} \right){{\phi }_{v}}\left( {{\rho }_{y}} \right),
\end{equation}
де $ {{\hat{V}}_{1}}\left( {{\rho }_{X}} \right)- $ нова операторнозначна польова функція, а $ {{\phi }_{v}}\left( {{\rho }_{y}} \right)- $ власна функція оператора \eqref{Hv_internal}, що відповідає власному значенню, яке позначимо ${\mu_{G}^{2}}/{2}\;$:
\begin{equation}\label{vlasni_znachenna_H_v_internal}
-\frac{1}{2}\sum\limits_{l=4}^{9}{\left( \frac{{{\partial }^{2}}{{\phi }_{v}}\left( {{\rho }_{y}} \right)}{\partial {{\left( \rho _{y}^{l} \right)}^{2}}} \right)+\frac{1}{2}m_{G}^{2}{{\phi }_{v}}\left( {{\rho }_{y}} \right)+3{{g}^{2}}\sqrt{\frac{6}{k}}\left( {{v}_{0}}\left( {{\rho }_{y}} \right) \right){{\phi }_{v}}\left( {{\rho }_{y}} \right)}=\frac{\mu_{G}^{2}}{2}{{\phi }_{v}}\left( {{\rho }_{y}} \right),
\end{equation}
отримаємо:
\begin{equation}\label{l_v_pisla_vlasnih_znachen}
\begin{split}
& l_{{{v}_{1}}}^{\left( 0 \right)}=\left( \frac{1}{2}\sum\limits_{{{l}_{1}}=0}^{3}{\sum\limits_{{{l}_{2}}=0}^{3}{{{g}^{{{l}_{1}}{{l}_{2}}}}\frac{\partial {{{\hat{V}}}_{1}}\left( {{\rho }_{X}} \right)}{\partial \rho _{X}^{{{l}_{1}}}}\frac{\partial {{{\hat{V}}}_{1}}\left( {{\rho }_{X}} \right)}{\partial \rho _{X}^{{{l}_{2}}}}}}-\frac{\mu_{G}^{2}}{2}{{\left( {{{\hat{V}}}_{1}}\left( {{\rho }_{X}} \right) \right)}^{2}} \right){{\left( {{\phi }_{y}}\left( {{\rho }_{y}} \right) \right)}^{2}}. \\ 
\end{split}
\end{equation}
Якщо власну функцію $ {{\phi }_{v}}\left( {{\rho }_{y}} \right) $ оператора $ \hat{H}_{v}^{\text{internal}}$ \eqref{Hv_internal} нормувати на одиницю, то підставляючи лагранжіан \eqref{l_v_pisla_vlasnih_znachen} у дію і інтегруючи по внутрішніх змінних $ {{\rho }_{y}}$ для поля $ {{\hat{V}}_{1}}\left( {{\rho }_{X}} \right) $, отримаємо звичайний лагранжіан дійсного скалярного поля, для якого процедура квантування призводить до операторів народження і знищення частинок із масою ${{\mu}_{G}}$. 

Як видно з \eqref{Rozcladanna_v_pramu_sumu}, поле $ {{\hat{V}}_{1}}\left( {{\rho }_{X}} \right) $ пов'язано з розкладом у пряму суму інваріантних підпросторів лінійного простору двоіндексних тензорів. Тому, оператори народження і знищення, які відповідають цьому полю, повинні змінювати числа заповнення двоглюонних станів. В той же час, оператор \eqref{Hv_internal} і його власна функція $ {{\phi }_{v}}\left( {{\rho }_{y}} \right) $ є тричастинковими. Тому оператор \eqref{Hv_internal} ми не можемо інтерпретувати як внутрішній гамільтоніан двоглюонного стану, а його власну функцію - як таку, що описує цей стан. Це можна пояснити тим, що ми розглядаємо як одночастинкові глюонні поля, введені при подовженні похідних \eqref{Kovariantni_pohidni_sz_3_polami}, так і двочастинкові, введені в \eqref{Lagrangian_z_dvogluonnimi_polami} як такі, що взаємодіють із внутрішнім станом кварків у баріоні. Така взаємодія може надати інформацію про результат вимірювання розташування трьох кварків у баріоні, але не може слугувати вимірюванням розташування двох глюонів усередині двочастинкового стану. Так відбувається тому, що як мінімум один глюон повинен провзаємодіяти одночасно як мінімум із двома кварками, які при вимірюванні можуть бути виявленими в двох різних точках. Тому, за таких умов, взаємодія з кварками не зможе перевести пару глюонів у стан власний для їх відносних координат. Внаслідок цього, власна функція ${{\phi }_{v}}\left( {{\rho }_{y}} \right) $ оператора \eqref{Hv_internal} не описує стан пари глюонів. Однак, оскільки двоглюонні поля взаємодіють із баріоном посередництвом взаємодії із кварками всередині нього, вершина взаємодії, стаючи нелокальною, повинна враховувати внутрішню структуру баріону. Переписуючи лагранжіан взаємодії \eqref{L_int_1} з урахуванням \eqref{Razlojenie_v1_ro} в виді: 
\begin{equation}\label{	l_int_cherez_verhinu}
	\begin{split}
		& l_{\operatorname{int}}^{1}=-\frac{{{g}^{2}}}{\left( 9{{\mu }_{q}} \right)}\sqrt{\frac{6}{k}}{{{\hat{\bar{\Psi }}}}_{{{s}_{1}}}}\left( {{\rho }_{X}} \right){{{\hat{\Psi }}}_{{{s}_{1}}}}\left( {{\rho }_{X}} \right){{{\hat{V}}}_{1}}\left( {{\rho }_{X}} \right){{\phi }^{*}}\left( {{\rho }_{y}} \right)\phi \left( {{\rho }_{y}} \right){{\phi }_{v}}\left( {{\rho }_{y}} \right)- \\ 
		& -{{g}^{2}}\sqrt{\frac{6}{k}}{{\left( {{{\hat{V}}}_{1}}\left( {{\rho }_{X}} \right) \right)}^{3}}{{\left( {{\phi }_{v}}\left( {{\rho }_{y}} \right) \right)}^{3}}, \\
	\end{split}
\end{equation} 
бачимо, що функція $ {{\phi }_{v}}\left( {{\rho }_{y}} \right) $ описує як вершину взаємодії протону з двоглюонним полем, так і вершину самодії двоглюонного поля, що взаємодіє із тричастинковим протонним полем. Зауважимо, що квадрат модуля функції $\phi \left( {{\rho }_{y}} \right),$ яка є власною функцією внутрішнього гамільтоніану \eqref{sam_vnutrihnij_hamiltonian}, і розглядається нами як координатна частина власного для енергії внутрішнього стану системи кварків у протоні, має ймовірнісний сенс. В той же час, вершина взаємодії в обраному нами представленні взаємодії, впливає на залежність від часу стану релятивістської квантової системи і не має прямого ймовірнісного сенсу. Тому нелокальна вершина не описуватиметься функцією $\phi \left( {{\rho }_{y}} \right),$ що пояснює появу в вершинах лагранжіану нової функції $ {{\phi }_{v}}\left( {{\rho }_{y}} \right). $ 
У той же час, як показано в роботах \cite{Korotca_statta_v_UJP, Ptashynskiy2019MultiparticleFO,Chudak:2016} двоглюонне поле можна розглянути незалежно від трикваркового, на основі міркувань, аналогічних тим, що були використані в цій роботі для трикваркового поля. При цьому, подібно до того, як у попередніх міркуваннях виник внутрішній гамільтоніан трикваркової системи, виникає внутрішній гамільтоніан двоглюонної системи і його власні функції вже можна інтерпретувати як такі, що характеризують власні стани двоглюонної системи. Відділяючи залежність польових функцій такого поля від координат Якобі, що відповідають центру мас, і залежність від внутрішніх змінних, можна досягти того, щоб лагранжіан операторнозначної польової функції від координат центру мас після інтегрування по внутрішнім змінним збігся з лагранжіаном 
\eqref{l_v_pisla_vlasnih_znachen} також після виконання інтегрування по внутрішніх змінних. Це дає змогу далі не розглядати два різних двоглюонних поля, а розглядати одне поле $ {{\hat{V}}_{1}}\left( {{\rho }_{X}} \right).$ Тоді це поле взаємодіятиме із протонами і мезонами 
\cite{Korotca_statta_v_UJP, Ptashynskiy2019MultiparticleFO,Chudak:2016} і самодіятиме по закону $ \sim {{\left( {{{\hat{V}}}_{1}}\left( {{\rho }_{X}} \right) \right)}^{3}}.$ Цим можна скористатися для опису процесів пружного і непружного розсіяння протонів.

\begin{figure}
	\centering
	\includegraphics[width=0.7\linewidth]{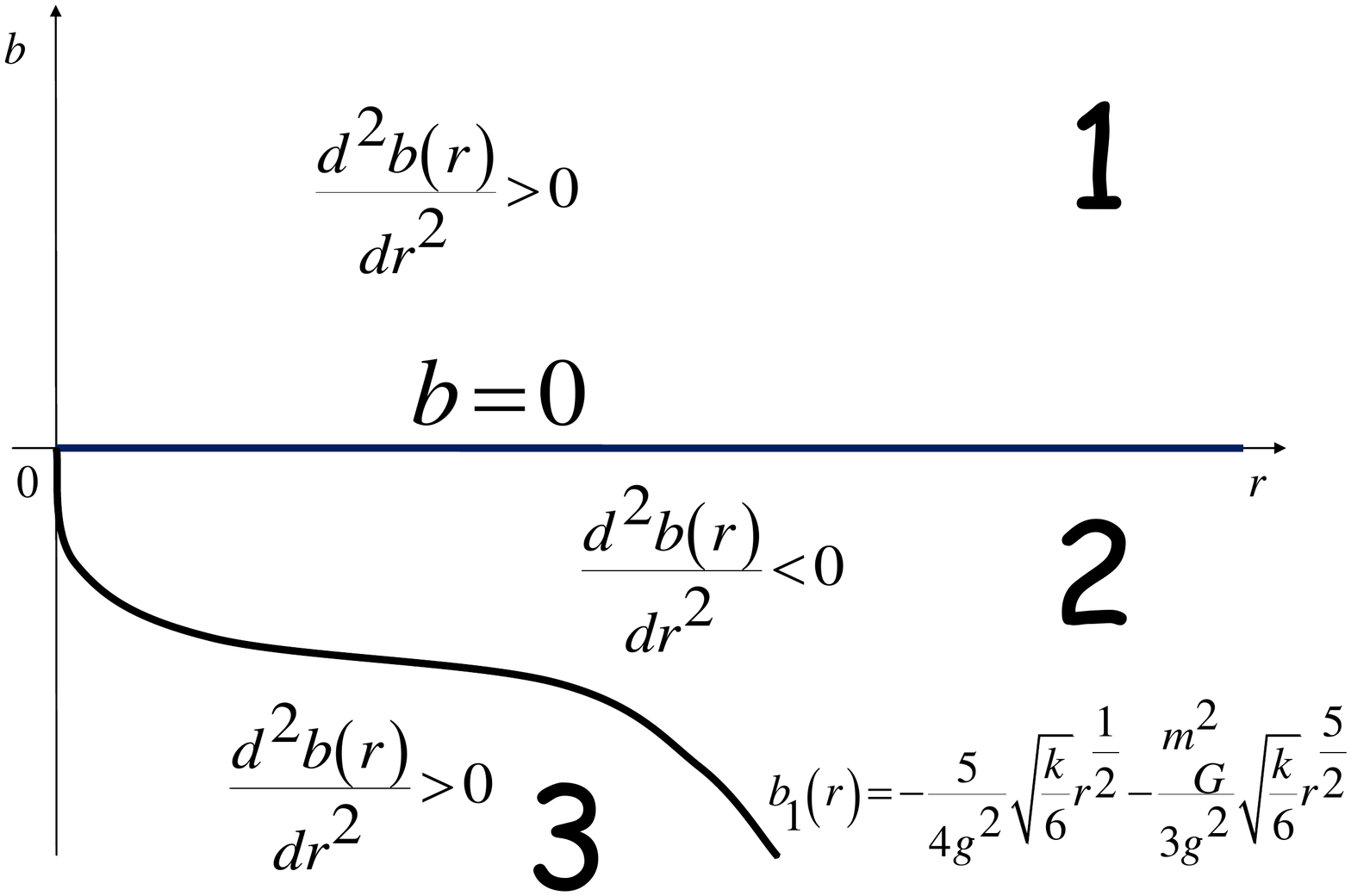}
	\caption[Рис.\ref{fig:oblastipriscorenna}]{Області постійного знаку другої похідної функції $b\left(r\right).$}
	\label{fig:oblastipriscorenna}
\end{figure}

У попередні співвідношення у виді потенційної енергії (з точністю до коефіцієнту) входить функція $ {{v}_{0}}\left( {{\rho }_{y}} \right),$ яка визначається рівнянням \eqref{rivnanna_dla_v0}.  Розглянемо сферично симетричний розв'язок цього рівняння, тобто такий розв'язок, в який незалежні змінні входять у комбінації: 
\begin{equation}\label{r}
	r=\sqrt{\sum\limits_{a=4}^{9}{{{\left( {{\rho }^{a}} \right)}^{2}}}}.
\end{equation}
Тоді рівняння \eqref{rivnanna_dla_v0} приймає вид:
\begin{equation}\label{sfericno_simetrichne_rivnanna_dla_v_0}
	\frac{{{d}^{2}}{{v}_{0}}\left( r \right)}{d{{r}^{2}}}+\frac{5}{r}\frac{d{{v}_{0}}\left( r \right)}{dr}-m_{G}^{2}{{v}_{0}}\left( r \right)-3{{g}^{2}}\sqrt{\frac{6}{k}}{{\left( {{v}_{0}}\left( r \right) \right)}^{2}}=0.
\end{equation}
Для аналізу властивостей його розв'язків зручно перейти до нової невідомої функції $ b\left( r \right) $: 
 \begin{equation}\label{v0_ravno_a_na_b}
 	{{v}_{0}}\left( r \right)=a\left( r \right)b\left( r \right),	
 \end{equation}
і обираючи функцію $ a\left( r \right) $ такою, щоб у рівнянні для $b\left( r \right)$ був відсутній доданок, що містить першу похідну. Виходячи з цієї вимоги покладемо:
\begin{equation}\label{v0_r_minus_pat_vtorix_na_b}
{{v}_{0}}\left( r \right)={{r}^{-\frac{5}{2}}}b\left( r \right).
\end{equation} 
Тоді для $b\left( r \right)$ отримаємо рівняння:
\begin{equation}\label{Rivnanna_dla_b_ot_r}
	\frac{{{d}^{2}}b\left( r \right)}{d{{r}^{2}}}=b\left( r \right)\left( \frac{15}{4}\frac{1}{{{r}^{2}}}+m_{G}^{2}+3{{g}^{2}}\sqrt{\frac{6}{k}}\frac{b\left( r \right)}{{{r}^{\frac{5}{2}}}} \right).
\end{equation}
При $r\to +0$, виходячи з \eqref{v0_r_minus_pat_vtorix_na_b} маємо або сингулярний розв'язок, або розв'язок $ b\left( r \right)$ повинен наближатися до нуля не менш швидко ніж $ {{r}^{\frac{5}{2}}}.$ Розглянемо регулярний у нулі розв'язок. Тоді при $r\to +0$ основний внесок у вираз в дужках у рівнянні \eqref{Rivnanna_dla_b_ot_r} вносить перший доданок $ \left( {15}/{4}\; \right)\left( {1}/{{{r}^{2}}}\; \right).$ Тому при $r\to +0$ рівняння \eqref{Rivnanna_dla_b_ot_r} наближається до
\begin{equation}\label{Rivnanna_dla_b_pri_r_nol}
	\frac{{{d}^{2}}b\left( r \right)}{d{{r}^{2}}}=\frac{15}{4}\frac{1}{{{r}^{2}}}b\left( r \right).
\end{equation}
Оскільки в обох частинах рівності \eqref{Rivnanna_dla_b_pri_r_nol} до функції $b\left( r \right)$ застосовуються такі операції, які понижують ступінь на 2 одиниці, то розв'язок цього рівняння можна знаходити в виді $b\left( r \right)={{r}^{n}}.$ Підставляючи це представлення в рівняння \eqref{Rivnanna_dla_b_pri_r_nol} отримаємо квадратне рівняння для $ n. $ Це рівняння має два розв'язки $ n={5}/{2}\;,n=-{3}/{2}\; $. Оскільки ми розглядаємо регулярний при $r\to +0$ розв'язок, то випадок $n=-{3}/{2}\;$ ми відкидаємо. Тоді маємо:
\begin{equation}\label{asimptotica_pri_r_nablijaetsa_do_nula}
	b\left( r \right)\xrightarrow{r\to +0}C{{r}^{\frac{5}{2}}},{{v}_{0}}\left( r \right)\xrightarrow{r\to +0}C.
\end{equation}
Задання певного значення константи $C$ визначає граничну умову, яка накладається на функцію $b\left( r \right)$ при $r\to +0$. Розглянемо поведінку розв'язків рівняння \eqref{Rivnanna_dla_b_ot_r} при різних значеннях константи $C.$ Для цього зазначимо, що якщо ми розглянемо координатну півплощину, у якої на вісі абсцис відкладається значення $r>0,$ а по вісі ординат - значення $b,$ то як видно з правої частини рівняння \eqref{Rivnanna_dla_b_ot_r} пряма 
\begin{equation}\label{prama}
b\left( r \right)=0,
\end{equation}
і крива 
\begin{equation}\label{kriva}
	b_{1}\left( r \right)=-\frac{5}{4{{g}^{2}}}\sqrt{\frac{k}{6}}{{r}^{\frac{1}{2}}}-\frac{m_{G}^{2}}{3{{g}^{2}}}\sqrt{\frac{k}{6}}{{r}^{\frac{5}{2}}},
\end{equation}
розбивають цю півплощину на області, показані на рис.\ref{fig:oblastipriscorenna}, всередині яких друга похідна $ {{{d}^{2}}b\left( r \right)}/{d{{r}^{2}}}\; $ має постійні знаки.
Розглянемо випадок $C>0.$ Тоді, як видно з \eqref{asimptotica_pri_r_nablijaetsa_do_nula} при  
 $r\to +0$ маємо $ b\left( r \right)>0 $ і графік функції $b\left( r \right)$ потрапляє в область $ 1 $ на рис.\ref{fig:oblastipriscorenna}. При цьому, перша похідна $ \frac{db\left( r \right)}{dr}=\frac{5}{2}C{{r}^{\frac{3}{2}}} $ в області малих $r$ додатна. Таким чином, маємо, що при малих $r$ і сама функція $b\left( r \right)$ і її перша і друга похідні є додатними. Але, оскільки, всюди в області $ 1 $ на рис.\ref{fig:oblastipriscorenna}, друга похідна додатна, то перша похідна повинна зростати і не може стати від'ємною. Тому, за умови $C>0$ матимемо монотонно зростаючий розв'язок, графік якого повністю розташований в області $ 1. $ Розглянемо, якою буде асимптотика цього розв'язку при $r\to +\infty. $ Оскільки нас цікавить поведінка потенційної енергії $ {{v}_{0}}\left( r \right)={{r}^{-{5}/{2}\;}}b\left( r \right), $ розглянемо асимптотику розв'язку $b\left( r \right)$ при різних припущеннях відносно $ \underset{r\to +\infty }{\mathop{\lim }}\,\left( {{r}^{-{5}/{2}\;}}b\left( r \right) \right).$ Припускаючи, що $ \underset{r\to +\infty }{\mathop{\lim }}\,\left( {{r}^{-{5}/{2}\;}}b\left( r \right) \right)=0 $ отримаємо, що рівняння \eqref{Rivnanna_dla_b_ot_r} при $r\to +\infty$ наближатиметься до 
 \begin{equation}\label{Rivnana_v_pripuchenni_nol}
 	\frac{{{d}^{2}}b\left( r \right)}{d{{r}^{2}}}=b\left( r \right)\left( \frac{15}{4}\frac{1}{{{r}^{2}}}+m_{G}^{2} \right).
 \end{equation} 
Оскільки $m_{G}^{2}$ довільний параметр, який з'явився після обезрозмірювання \eqref{Bezrozmirni_masi} відповідного параметру в лагранжіані \eqref{Perehid_vid_q_do_z}, ми не знаємо чому він дорівнює і тому розглянемо два випадки $ m_{G}^{2}=0$ і $ m_{G}^{2}>0.$ У першому випадку рівняння \eqref{Rivnana_v_pripuchenni_nol} зведеться до рівняння \eqref{Rivnanna_dla_b_pri_r_nol}. Його два лінійно незалежні розв'язки $ {{r}^{{5}/{2}\;}}$ і $ {{r}^{-{3}/{2}\;}}$. Але, оскільки $b\left( r \right)$ у нашому випадку монотонно зростаюча функція, асимптотика не може зводитись до $ {{r}^{-{3}/{2}\;}}.$ Тому, маємо $ b\left( r \right)\xrightarrow{r\to +\infty }{{r}^{{5}/{2}\;}},$ у випадку $ m_{G}^{2}=0.$ У випадку $ m_{G}^{2}>0$ у рівнянні \eqref{Rivnana_v_pripuchenni_nol} при великих $r$ першим доданком у дужках можна знехтувати у порівнянні із другим. Тоді асимптотика монотонно зростаючого розв'язку буде $ b\left( r \right)\xrightarrow{r\to +\infty }\exp \left( \left| {{m}_{G}} \right|r \right). $ Але в обох випадках ми прийшли до протиріччя із вихідним припущенням   
$ \underset{r\to +\infty }{\mathop{\lim }}\,\left( {{r}^{-{5}/{2}\;}}b\left( r \right) \right)=0,$ що доводить невірність цього припущення.

Припустимо тепер $ \underset{r\to +\infty }{\mathop{\lim }}\,\left( {{r}^{-{5}/{2}\;}}b\left( r \right) \right)=K,$ де $ K $ - деяка константа, яка внаслідок додатності в усіх точках розв'язку $b\left( r \right)  $ при обраній граничній умові, також повинна бути додатною. Тоді матимемо асимптотику монотонно зростаючого розв'язку $ b\left( r \right)\xrightarrow{r\to +\infty }\exp \left( r\sqrt{m_{G}^{2}+K} \right), $ яка також призведе до $ \underset{r\to +\infty }{\mathop{lim}}\,\left( {{r}^{{5}/{2}\;}}b\left( r \right) \right)=+\infty, $ що протирічить вихідному припущенню $ \underset{r\to +\infty }{\mathop{\lim }}\,\left( {{r}^{-{5}/{2}\;}}b\left( r \right) \right)=K.$ 

Розглянемо тепер припущення $ \underset{r\to +\infty }{\mathop{\lim }}\,\left( {{r}^{-{5}/{2}\;}}b\left( r \right) \right)=+\infty.$ Тоді рівняння \eqref{Rivnanna_dla_b_ot_r} наближається при $r\to +\infty $ до рівняння
\begin{equation}\label{Rivnanna_pri_nescinchenosti}
	\frac{{{d}^{2}}b\left( r \right)}{d{{r}^{2}}}=3{{g}^{2}}\sqrt{\frac{6}{k}}\frac{{{\left( b\left( r \right) \right)}^{2}}}{{{r}^{\frac{5}{2}}}}.
\end{equation}
Розглянемо функцію: 
\begin{equation}\label{Pribliznij_rozvazok_na_nescinchenosti }
	f\left( r \right)=\exp \left( \int\limits_{{{r}_{0}}}^{r}{\sqrt{3{{g}^{2}}\sqrt{\frac{6}{k}}\frac{{{\left( b\left( {{r}_{1}} \right) \right)}^{2}}}{{{r}^{\frac{5}{2}}}}}d{{r}_{1}}} \right).
\end{equation} 
Тут $ {r}_{0} -$довільне скінчене значення змінної $r.$ Для цієї функції маємо:
\begin{equation}\label{Druga_pohidna}
	\begin{split}
  & \frac{{{d}^{2}}}{d{{r}^{2}}}\left( f\left( r \right) \right)=3{{g}^{2}}\sqrt{\frac{6}{k}}\frac{{{\left( b\left( r \right) \right)}^{2}}}{{{r}^{\frac{5}{2}}}}f\left( r \right)+ \\ 
& +\frac{1}{2}\frac{d}{dr}\left( \ln \left( \sqrt{3{{g}^{2}}\sqrt{\frac{6}{k}}\frac{{{\left( b\left( r \right) \right)}^{2}}}{{{r}^{\frac{5}{2}}}}} \right) \right)f\left( r \right). \\ 		
	\end{split}
\end{equation} 
Оскільки функція під логарифмом за припущенням наближається до нескінченності, то її логарифм змінюється повільно і другим доданком у рівнянні \eqref{Druga_pohidna} можна знехтувати, у порівнянні із першим. Тоді функція \eqref{Pribliznij_rozvazok_na_nescinchenosti } може розглядатися як наближений розв'язок рівняння \eqref{Rivnanna_dla_b_ot_r} при $r\to +\infty. $ Оскільки перша похідна від функції \eqref{Pribliznij_rozvazok_na_nescinchenosti } є додатною, то цей розв'язок, як і потрібно, буде монотонно зростаючим.  Оскільки за припущенням підінтегральний вираз у показнику експоненти наближається при $r\to +\infty $
до нескінченності, то й сам розв'язок також наближається до нескінченності. Наявність експоненти забезпечує значно більшу швидкість наближення до нескінченності ніж ${{r}^{{5}/{2}\;}},$ що узгоджується із зробленим припущенням $ \underset{r\to +\infty }{\mathop{\lim }}\,\left( {{r}^{-{5}/{2}\;}}b\left( r \right) \right)=+\infty.$ Таким чином, єдине припущення, яке не призвело до протиріччя, призводить до висновку, що потенціал взаємодії кварків у протоні наближається до нескінченності із зростанням відстані між кварками. Цей висновок підтверджується результатом чисельного розрахунку, який показаний на рис.\ref{fig:v0otr}:

\begin{figure}
	\centering
	\includegraphics[width=0.7\linewidth]{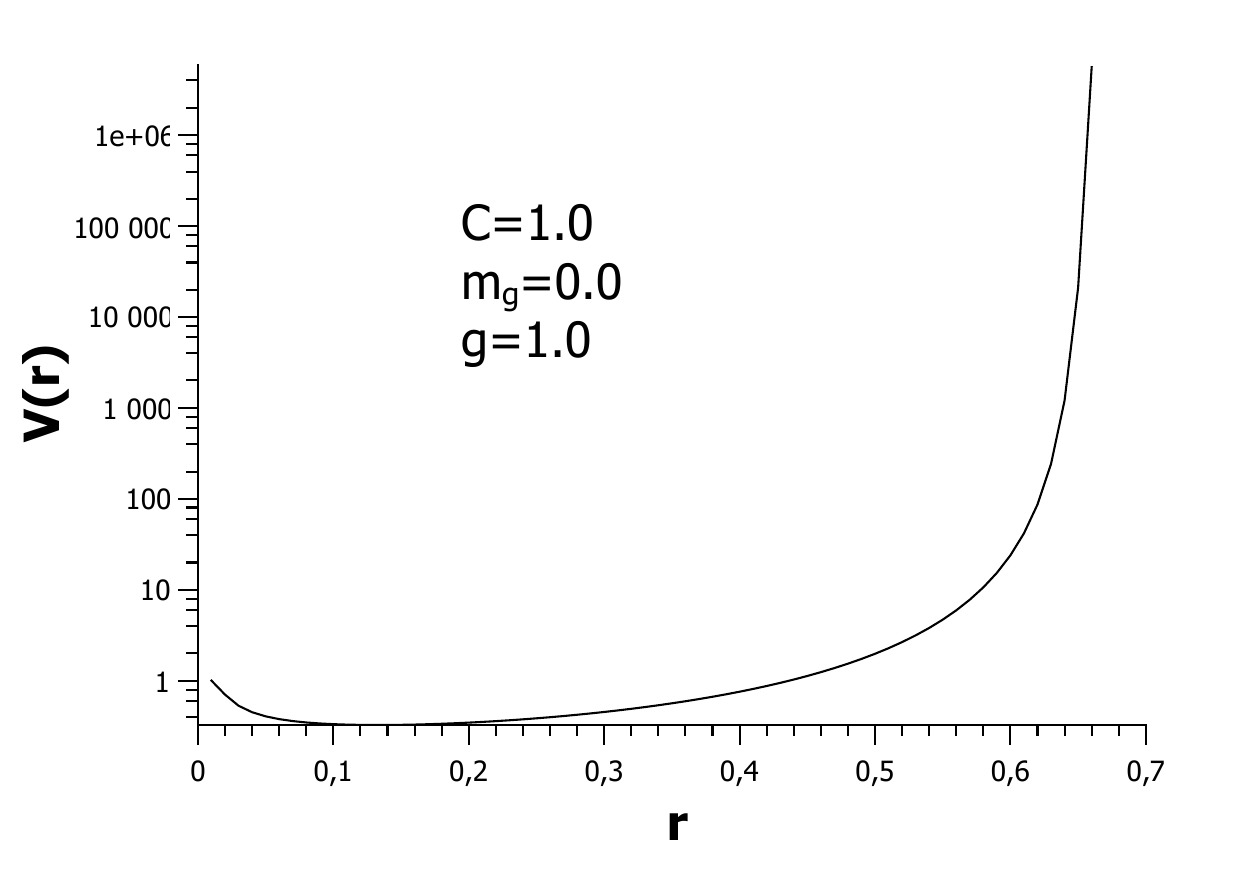}
	\caption[Рис.\ref{fig:v0otr}]{Залежність $v_{0}\left(r\right)$, отримана шляхом чисельного розв'язку рівняння \eqref{Rivnanna_dla_b_ot_r} і наступного перетворення \eqref{v0_r_minus_pat_vtorix_na_b} при $ C=1.0,g=1.0,m_{g}=0.$ По вісі ординат використано логарифмічний масштаб.}
	\label{fig:v0otr}
\end{figure}

Розрахунок при $ m_{G}^{2}\ne 0$ призводить до подібного результату, показаного на рис.\ref{fig:votrmg10}:
\begin{figure}
	\centering
	\includegraphics[width=0.7\linewidth]{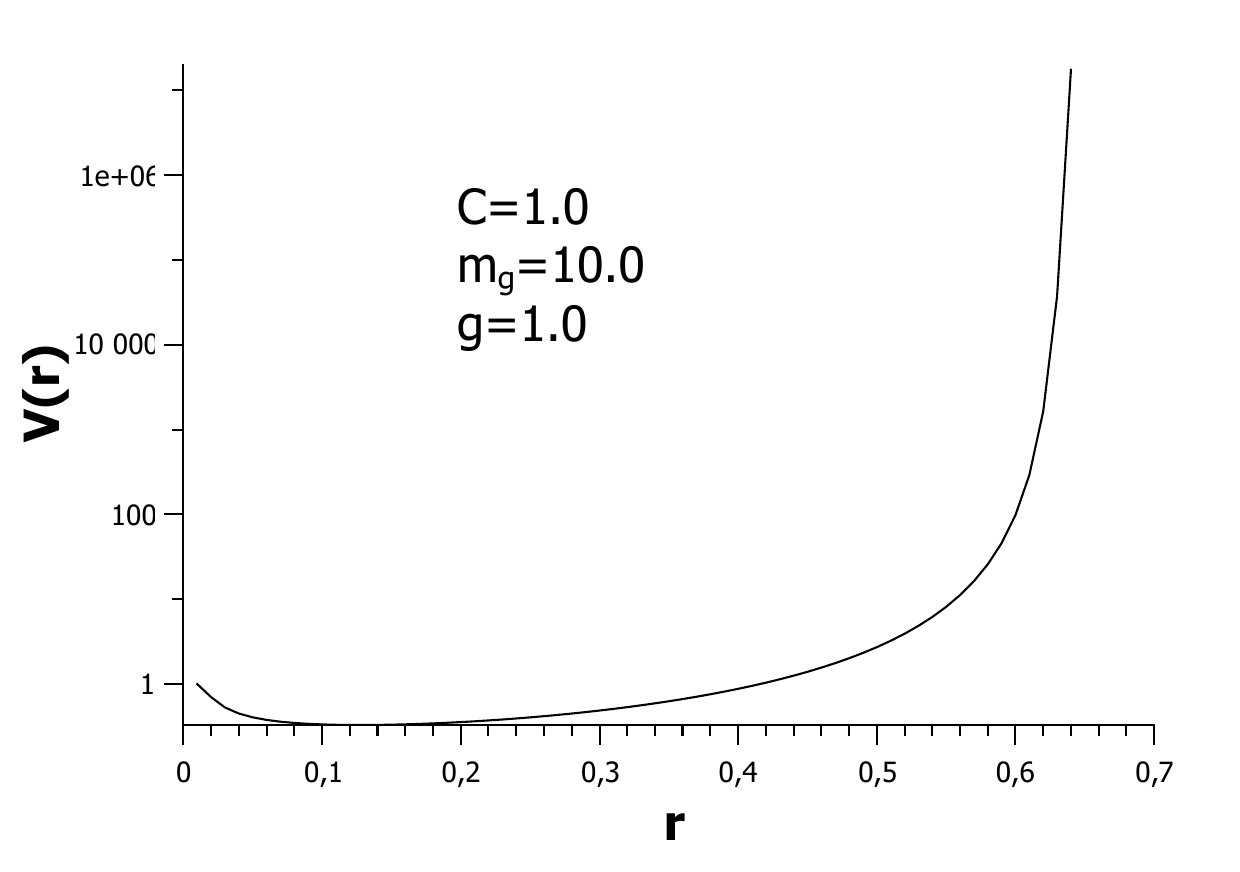}
	\caption[Рис.\ref{fig:votrmg10}]{Залежність $v_{0}\left(r\right)$, отримана шляхом чисельного розв'язку рівняння \eqref{Rivnanna_dla_b_ot_r} і наступного перетворення \eqref{v0_r_minus_pat_vtorix_na_b} при $ C=1.0,g=1.0,m_{g}=10.$ По вісі ординат використано логарифмічний масштаб.}
	\label{fig:votrmg10}
\end{figure}
 
Ці результати узгоджуються із тим фактом, що кварки в протоні знаходяться у стані конфайнменту. 

У випадку $C<0$ залежність $b\left( r \right)$ при малих $ r $ потрапляє в область 2 на Рис. \ref{fig:oblastipriscorenna}. В цьому випадку, аналіз знаків першої і другої похідної приводить до висновку, що залежність $b\left( r \right)$, яка коливатиметься навколо графіка $b_{1}\left( r \right)$ на Рис. \ref{fig:oblastipriscorenna}. Результат відповідного чисельного розрахунку показаний на Рис. \ref{fig:botrcminus0i5}: 
\begin{figure}
	\centering
	\includegraphics[width=0.7\linewidth]{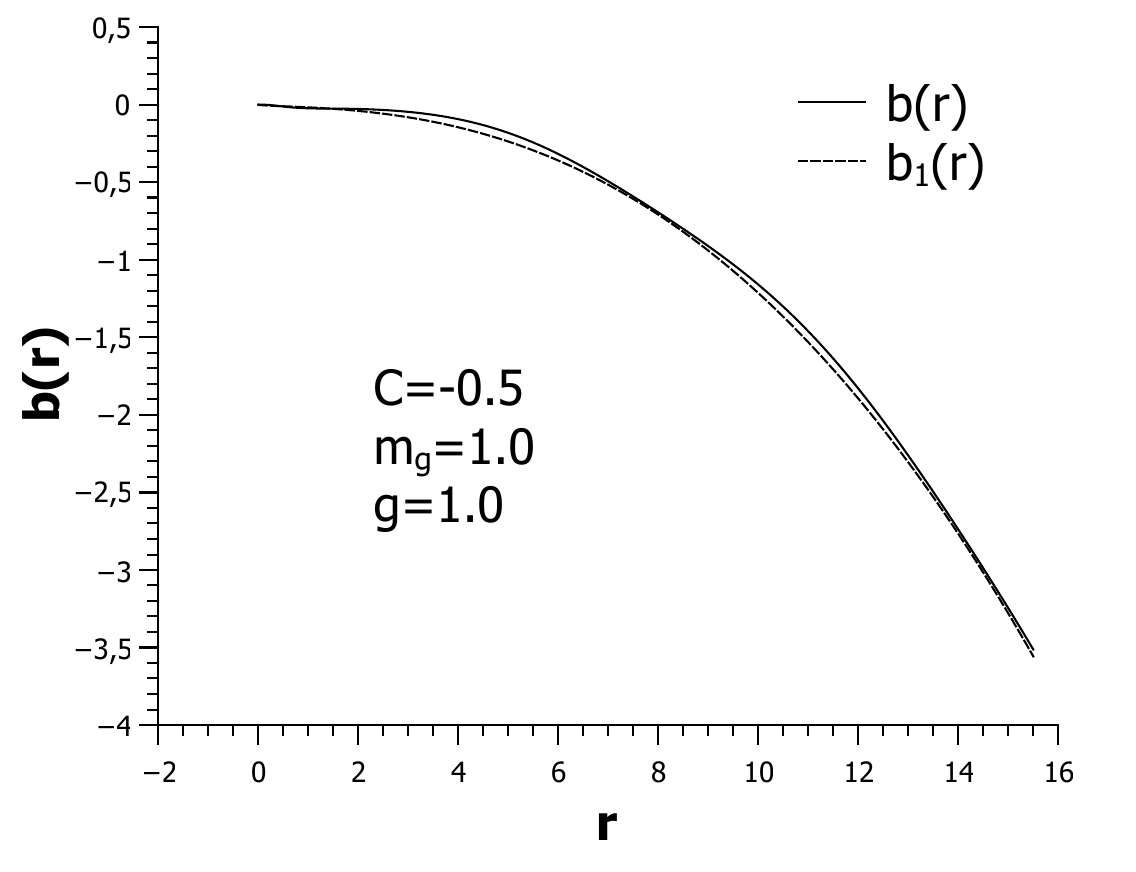}
	\caption{Залежність $b\left( r \right)$, отримана в результаті чисельного розрахунку при $C=-0.5,m_{g}=1.0,g=1.0.$}
	\label{fig:botrcminus0i5}
\end{figure}
Відповідний потенціал $V\left( r \right)$ показано на Рис.\ref{fig:vorrcminus0i5}:
\begin{figure}
	\centering
	\includegraphics[width=0.7\linewidth]{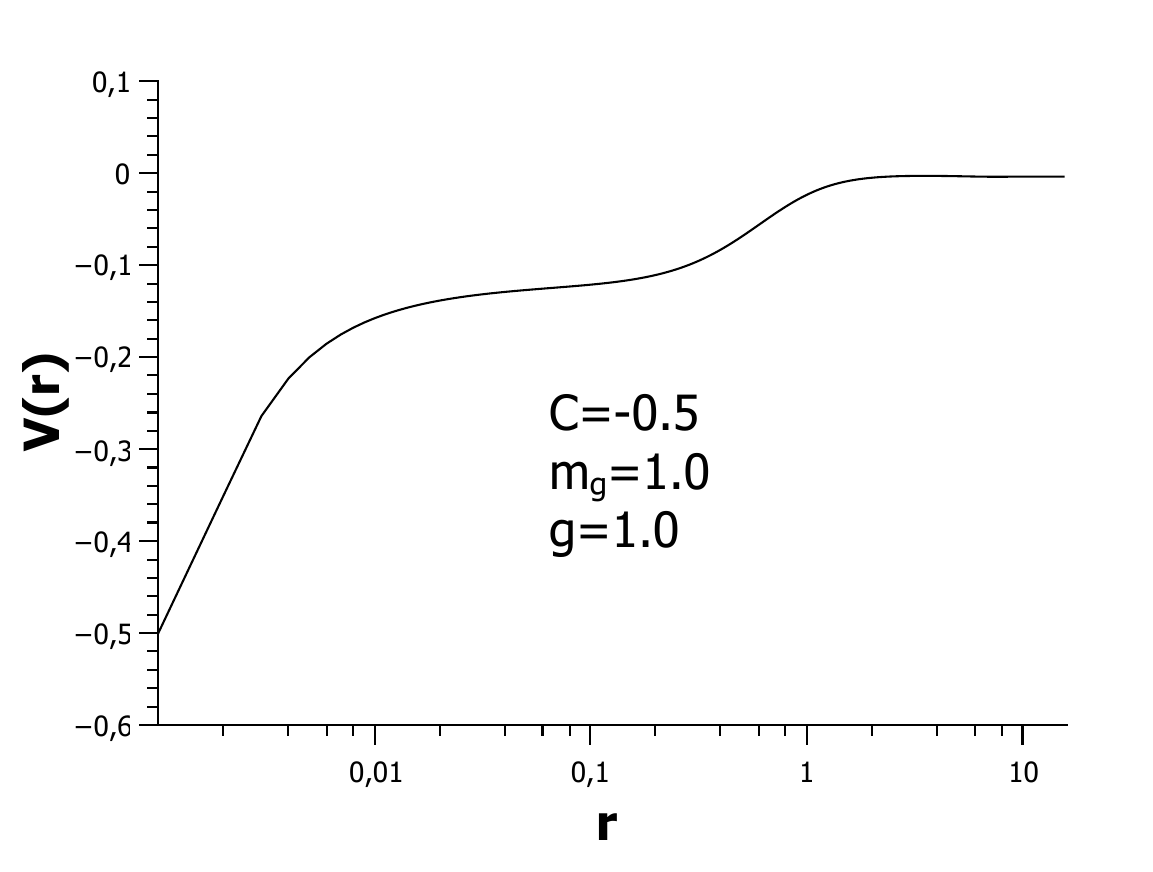}
	\caption{Залежність $V(r),$ отримана в результаті чисельного розрахунку при $C=-0.5,m_{g}=1.0,g=1.0.$. По вісі абсцис використано логарифмічний масштаб}
	\label{fig:vorrcminus0i5}
\end{figure}
Як видно з Рис.\ref{fig:vorrcminus0i5} за такої потенційної енергії може існувати зв'язаний стан трьох ферміонів, але він матиме скінчену енергію зв'язку. Тому для нашої задачі опису адрону як трикваркової системи випадок $C<0$ не представляє інтересу. Можлива фізична інтерпретація цього випадку обговорюється у наступному розділі. Аналогічний випадок для зв'язаного стану двох калібрувальних бозонів, як розглядалося в роботах \cite{Korotca_statta_v_UJP, Ptashynskiy2019MultiparticleFO,HiggsJPS} описує механізм спонтанного порушення симетрії.
 
\section{Обговорення результатів і висновки}
Запропонована в цій і попередніх роботах модель багаточастинкових полів може бути використана для спроби опису експериментів з пружного і непружного розсіяння адронів. Для протон-протонних зіткнень ми маємо тричастинкове біспінорне поле, що відповідає протонам і антипротонам і це поле взаємодіє із двочастинковим глюбольним полем. Це глюбольне поле самодіє внаслідок самодії неабелева калібрувального поля і може взаємодіяти із двочастинковим мезонним полем, або з іншими тричастинковими полями. З цих складових частин можна побудувати діаграми, які відповідатимуть пружним і непружним процесам. При цьому, непертурбативні ефекти описуватимуться внутрішніми гамільтоніанами багаточастинкових полів. Те, що такий опис виявився нерелятивістським можна пояснити на основі аргументів, викладених у \cite{Ptashynskiy2019MultiparticleFO}. Суть цих аргументів полягає в тому, що квантова механіка - принципово нелокальна теорія. Якщо розглядати вимірювання в координатному представленні, то процес взаємодії із вимірювачем повинен бути реалізований таким чином, щоб частинки системи могли провзаємодіяти із приладом у будь-якій точці деякої області. Тобто, прилад не є локалізований у якійсь точці, а розподілений по області. Внаслідок цього, коли прилад взаємодіє з частинками багаточастинкової системи в деякій області, зміна стану такої системи відбувається миттєво, бо кожна частинка може провзаємодіяти з приладом у будь-якій точці, і тому зміна стану не потребує розповсюдження взаємодії з однієї точки простору-часу до іншої. Відповідно скінченність швидкості цього розповсюдження перестає, в такому випадку, бути суттєвою і тому опис стає аналогічним нерелятивістському. 

В той же час, в цій роботі ми розглянули найбільш простий варіант багаточастинкової моделі, маючи на меті отримати опис найбільш близький до звичайної одночастинкової теорії. Тому, при обговоренні сенсу квантованого багаточастинкового поля, ми припустили, що залежність від координат центру мас відділяється від залежності від внутрішніх змінних. Це припущення не є критично важливим для квантування багаточастинкового поля. Дійсно, для надання польовим операторам сенсу операторів народження і знищення суттєвим є лише закон їх перетворення при просторово-часовому зсуві \cite{Bogolyubov_rus}. Але таке перетворення не зачіпає внутрішніх змінних і змінює лише просторово-часові координати центру мас. Тому, як саме внутрішні змінні входять у залежність польового оператору від його аргументів для процедури квантування не грає суттєвої ролі. Тому можна спробувати отримати інші розв'язки розглянутих в цій роботі динамічних рівнянь і їх фізично інтерпретувати. Зокрема, найбільш очевидним узагальненням розглянутої в роботі моделі є використання розкладу багаточастинкових операторів по власних станах внутрішнього гамільтоніану, а не обмежуватися внеском лише основного стану, як ми це зробили в цій роботі. Також представляє інтерес, розглянути моделі із зміною в процесі кількості частинок у внутрішніх станах взаємодіючих адронів і взаємодію багаточастинкових полів з одночастинковими. Однак, поки що незрозуміло, яким чином ввести відповідні доданки у внутрішні гамільтоніани, виходячи з калібрувального принципу введення взаємодій. 

Фактично, можна сказати що проблема, яка обговорювалася в цій роботі є наслідком протиріччя між нелокальним характером квантової механіки і локальним видом квантової теорії поля. Можливо це протиріччя проявляється в відомому факті розбіжності інтегралів, що відповідають деяким діаграмам Р. Фейнмана. Як відомо \cite{Bogolyubov_rus}, ці розбіжності виникають внаслідок невизначеності хронологічного спарювання польових операторів при співпадінні часових аргументів, тобто саме на підмножині одночасності, яка розглядалася вище. Ці розбіжності виникають на світловому конусі, тобто на підмножині точок тензорного добутку двох просторів Мінковського, яка відділяє область цього добутку, яка має спільні точки з підмножиною одночасності, від області, яка таких точок не має. З іншого боку, ці розбіжності виникають для деяких діаграм, які містять петлі, тобто багаточастинкові проміжні стани. Однак для таких станів, на підмножині одночасності можлива інша динаміка, ніж та, що описується за допомогою одночастинкових функцій Гріна. Можливо, в цьому випадку, стануть у нагоді розв'язки багаточастинкових рівнянь, які розглядалися в попередньому розділі і не призводять до конфайнменту. Тоді в багаточастинковій задачі матимемо суцільний спектр і стани цього спектру мали б враховуватися при інтегруванні по проміжним багаточастинковим станам.

%

\bibliographystyle{unsrtnat}

\bibliography{references-utf8}

\end{document}